\documentclass{amsart}

\usepackage{slashed}
\usepackage{adjustbox}
\usepackage{amsmath}	
\usepackage{amssymb}
\usepackage{graphicx}
\usepackage{mathrsfs}	
\usepackage{multicol}
\usepackage{soul}	
\usepackage{color}	
\usepackage{appendix}	
\usepackage{eufrak}
\usepackage[labelformat=simple]{subcaption}
\usepackage[backend=bibtex,style=numeric,citestyle=numeric,maxbibnames=99,giveninits,doi=false,isbn=false,url=false,eprint=false]{biblatex}

\usepackage{hyperref}	

\renewbibmacro{in:}{\ifentrytype{article}{}{\printtext{\bibstring{in}\intitlepunct}}}
 
\newcommand{\de}{\ensuremath{\partial}}
\newcommand{\dee}{\ensuremath{\textrm{d}}}

\newcommand{\inty}[4]{\ensuremath{ \int_{#1}^{#2} \! #3 \, \dee#4 }}

\newcommand{\field}[1]{\mathbb{#1}}
\newcommand{\ip}[2]{\ensuremath{ \left< \left. #1 \right| #2 \right> } }

\newtheorem*{theorem*}{Theorem}
\newtheorem*{problemstatement*}{Problem Statement}
\newtheorem*{hypothesis*}{Hypothesis}
\bibliography{../../../document_template/bibliography.bib}

\numberwithin{lemma}{section}
\numberwithin{example}{section}
\numberwithin{figure}{section}
\numberwithin{proposition}{section}
\numberwithin{equation}{section}
\numberwithin{theorem}{section}
\numberwithin{remark}{section}
\numberwithin{definition}{section}
\numberwithin{assumption}{section}

\allowdisplaybreaks

\title[Wave-packet propagation and the spectral localizer index]{Wave-packet propagation in a finite topological insulator and the spectral localizer index}

\author{Jonathan Michala, Alexander Pierson, Terry A. Loring, and Alexander B. Watson}

\begin{document}




\maketitle

\begin{abstract}
We consider a model of electrons in a finite topological insulator. We numerically study the propagation of electronic wave-packets localized near edges of the structure in the presence of defects and random disorder. We compare the propagation with computations of the \emph{spectral localizer index}: a spatially local topological index. We find that without disorder, wave-packets propagate along boundaries between regions of differing spectral localizer index with minimal loss, even in the presence of strong defects. With disorder, wave-packets still propagate along boundaries between regions of differing localizer index, but lose significant mass as they propagate. We also find that with disorder, the \emph{localizer gap}, a measure of the localizer index ``strength'', is generally smaller away from the boundary than without disorder. Based on this result, we conjecture that wave-packets propagating along boundaries between regions of differing spectral localizer index do not lose significant mass whenever the localizer gap is sufficiently large on both sides of the boundary.
\end{abstract}

\section{Introduction} \label{sec:intro}

\subsection{Motivation} 

Topological insulators are materials which conduct robustly at their physical edge \cite{2010HasanKane,BernevigHughes}. The robustness of this edge conductance has attracted huge attention for industrial applications because it may make possible novel low-power electronic devices, for example the recently proposed ``topological transistor'' \cite{2018Collinsetal}.

The robust edge conductance of topological insulators can be seen already at the level of solutions of a single-electron Schr\"odinger equation. Specifically, one observes that wave-packets formed from edge states (eigenfunctions of the system Hamiltonian localized near to the edge of the material) of topological insulators are minimally disrupted by defects as they propagate around the edge of the material. This remarkable wave phenomenon has by now been experimentally observed across many model wave systems governed by the Schr\"odinger equation \cite{2007Konigetal,2008HsiehQianWrayXiaHorCavaHasan,2009WangChongJoannopoulosSoljacic,2013Changetal,2013Rechtsmanetal,2015SusstrunkHuber,2017DelplaceMarstonVenaille}.

In recent years, significant mathematical progress has been made towards understanding this phenomenon. A particular focus of research has been the ``bulk-boundary correspondence'', which states that the existence of edge states is equivalent to non-triviality of topological invariants associated with the ``bulk'' (i.e. far from edges) material \cite{1993Hatsugai,1999Schulz-BaldesKellendonkRichter,2002KellendonkRichterSchulz-Baldes,KELLENDONK2004388,2004KellendonkSchulz-Baldes,2005ElgartGrafSchenker,2013GrafPorta,2015Loring,2016ProdanSchulz-Baldes,Kubota:2017aa,2018GrafShapiro,Graf2018,2018Bal,2019Drouot}. The fact of this link explains the observed persistence of edge states under strong perturbations, and explains to some extent the stable propagation of wave-packets formed from these edge states. However, relatively little attention has been paid to understanding the stability of edge wave-packet propagation directly (although \cite{2018Bal,2019Drouot} do address local details of ``topologically-protected'' propagation at edges).

\subsection{Aim of this paper}

The aim of this paper is to understand to what extent wave-packet propagation at edges of topological insulators can be predicted using the ``localizer index'', a \emph{spatially local} topological index proposed by Loring \cite{2015Loring}. More specifically, we are interested in whether wave-packets propagate stably along the boundary between regions of distinct localizer index, \emph{even when the material has strong defects and disorder}. We will focus on the ``$p_x + i p_y$'' model, a simple model of a two-dimensional time-reversal symmetry-broken (Chern) topological insulator which can be defined easily on arbitrary arrangements of points in $\field{R}^2$ \cite{2016FulgaPikulinLoring}. We describe our aims in more detail in Section \ref{sec:prob_stat}. 

\subsection{Summary of results}

Our results, stated roughly, can be summarized as follows. 
\begin{itemize}
\item In the absence of random disorder, wave-packets initialized at boundaries between regions of distinct localizer index propagate stably along the boundary, even when the edge has strong defects such as sites missing from the edge.
\item When the structure has significant disorder, wave-packets initialized at boundaries between regions of distinct localizer index may or may not propagate stably between regions of distinct localizer index. Hence index information alone does not provide useful information in this case. However, if this information is supplemented with secondary information about the ``strength'' of the index at each point, the combined information is somewhat predictive in determining the stability of propagating wave-packets. 
\end{itemize}
These results will be clarified and expanded on in Sections \ref{sec:ind_comp} and \ref{sec:propagation}. 

\subsection{Related work} Other than Loring \cite{2015Loring} (see also \cite{2010HastingsLoring,2011HastingsLoring,2017LoringSchulz-Baldes,2019Loring_2,2019Loring}), several authors have proposed local topological indices including Kitaev \cite{KITAEV20062}, and Bianco and Resta \cite{2011BiancoResta} (see also \cite{2010Prodan,2019MarcelliMonacoMoscolariPanati}). A recent work has studied the localizer on a similar model to that considered here, although without considering dynamics of solutions of the time-dependent Schr\"odinger equation \cite{2019Lozano-ViescaSchoberSchulz-Baldes}. Mitchell, Irvine, et al. have investigated topological insulators built from random point sets using Kitaev's local index and demonstrated numerically that wave-packets propagate robustly at the edge of such systems \cite{2018Mitchelletal}. 

\subsection{Structure of paper} The structure of this paper is as follows. In Section \ref{sec:model}, we review the ``$p_x + i p_y$ model'', the model of a two-dimensional time-reversal symmetry-broken topological insulator we will use for our numerical experiments. In Section \ref{sec:index}, we review the definitions of the localizer and localizer index in general. In Section \ref{sec:BB_correspondence}, we explain how the localizer and localizer index define a form of bulk-boundary correspondence. In Section \ref{sec:prob_stat} we define the aims of our project precisely. In Section \ref{sec:ind_comp}, we present our first results: computations of the localizer index and the localizer index ``strength'' (localizer gap). In Section \ref{sec:propagation}, we present our main results: comparisons of computations of the localizer index and gap with computations of wave-packet dynamics. 

\subsection*{Acknowledgements} This project started as an undergraduate research project by J. M. and A. P. lead by A. W. at Duke University. The project was first part of the DoMath program (Summer 2018), and then continued as a Research Independent Study (Fall 2018). The authors are grateful to the Duke Mathematics Department, especially Lenhard Ng and Heekyoung Hahn, for organizing the DoMath program. A. W. would like to acknowledge helpful conversations with Jianfeng Lu. 

\section{The $p_x + i p_y$ model} \label{sec:model}

We consider the $p_x + i p_y$ model as introduced by \cite{2016FulgaPikulinLoring}. We first describe this model on an infinite two-dimensional square lattice, where the model's Chern numbers can be easily defined and numerically computed (Section \ref{sec:mod_square}). We then describe the generalization of this model we will study in the remainder of this work, which is defined on arbitrary finite point sets of $\field{R}^2$. Working with this generalization will allow us to study the effects of quite general disorder (Section \ref{sec:mod_general}).

\subsection{$p_x + i p_y$ model on an infinite two-dimensional square lattice} \label{sec:mod_square} We consider the Hilbert space $\mathcal{H} := l^2\left(\field{Z}^2;\field{C}^2\right)$, denoting elements of this space by $\psi = \left( \psi_{m,n} \right)_{m,n \in \field{Z}^2}$ where $\psi_{m,n} \in \field{C}^2$. 
The Hamiltonian acts by
\begin{equation} \label{eq:struc}
\begin{split}
	\left( H \psi \right)_{m,n} = \phantom{=} &A^* \psi_{m-1,n} + B^* \psi_{m,n-1}     \\
        &\phantom{blablabla}+ V \psi_{m,n} + A \psi_{m+1,n} + B \psi_{m,n+1}, 
\end{split}
\end{equation}
where
\begin{equation}
        A := - t \sigma_3 - \frac{i \Delta}{2} \sigma_1 
        , \quad B := - t \sigma_3 - \frac{i \Delta}{2} \sigma_2 
        , \quad V := - \mu \sigma_3. 
\end{equation}
Here $\mu, t, \Delta$ are real parameters,
\begin{equation}
	\sigma_1 = \begin{pmatrix} 0 & 1 \\ 1 & 0 \end{pmatrix}, \quad \sigma_2 = \begin{pmatrix} 0 & - i \\ i & 0 \end{pmatrix}, \quad \sigma_3 = \begin{pmatrix} 1 & 0 \\ 0 & -1 \end{pmatrix},
\end{equation}
denote the Pauli matrices, and $M^*$ denotes the conjugate transpose (adjoint) of $M$.

Periodicity of the Hamiltonian implies eigenvectors $\Phi$ of $H$ can be chosen to satisfy the Bloch conditions $\Phi_{m+1,n} = e^{i k_1} \Phi_{m,n}, \Phi_{m,n+1} = e^{i k_2} \Phi_{m,n}$, where $k_1, k_2 \in [0,2\pi)^2$ are the quasi-momenta. For such eigenvectors $\chi := \Phi_{00}$ satisfies the $k$-dependent eigenvalue problem 
\begin{equation}
\begin{split}
    &H(k_1,k_2) \chi = E \chi   \\
    &H(k_1,k_2) := \left( - \mu - 2 t ( \cos(k_1) + \cos(k_2) ) \right) \sigma_3 + \Delta \sin(k_1) \sigma_1 + \Delta \sin(k_2) \sigma_2.
\end{split}
\end{equation}
The eigenvalues of this matrix, the Bloch band functions, are
\begin{equation}
    E_\pm(k_1,k_2) = \pm 2 t \sqrt{ (\Delta')^2 \sin^2(k_1) + (\Delta')^2 \sin^2 (k_2) + ( \mu' + \cos(k_1) + \cos(k_2) )^2 },
\end{equation}
where $\Delta' := \frac{ \Delta }{ 2 t }$, and $\mu' := \frac{ \mu }{ 2 t }$. 

Ignoring the trivial case where $t = 0$, for band touchings (and hence topological transitions) to occur it must be that
\begin{equation}
\begin{split}
    &( \Delta' )^2 \sin^2(k_1) = 0  \\
    &( \Delta' )^2 \sin^2(k_2) = 0  \\
    &\mu' + \cos(k_1) + \cos(k_2) = 0.
\end{split}
\end{equation}
There are two possibilities: $\Delta' \neq 0$ or $\Delta' = 0$, which we will consider in turn.

If $\Delta' \neq 0$, the first two conditions imply that band touchings can only occur when $k_1, k_2 \in \{0,\pi\}$. For the third condition to hold there are four possibilities. We require either $k_1 = k_2 = 0$ and $\mu' = - 2$, $k_1 = k_2 = \pi$ and $\mu' = 2$, $k_1 = \pi$, $k_2 = 0$ and $\mu' = 0$, or $k_1 = 0$, $k_2 = \pi$ and $\mu' = 0$. 

If $\Delta' = 0$, then band touchings cannot occur for $|\mu'| > 2$, but occur for every $|\mu'| \leq 2$. To see this, note that in this case we can always choose $k_1$ such that $|\mu' + \cos(k_1)| \leq 1$, and hence for such a $k_1$ we can always find $k_2$ such that
\begin{equation}
    \mu' + \cos(k_1) + \cos(k_2) = 0.
\end{equation}

\begin{figure}
\includegraphics[scale=.5,draft=false]{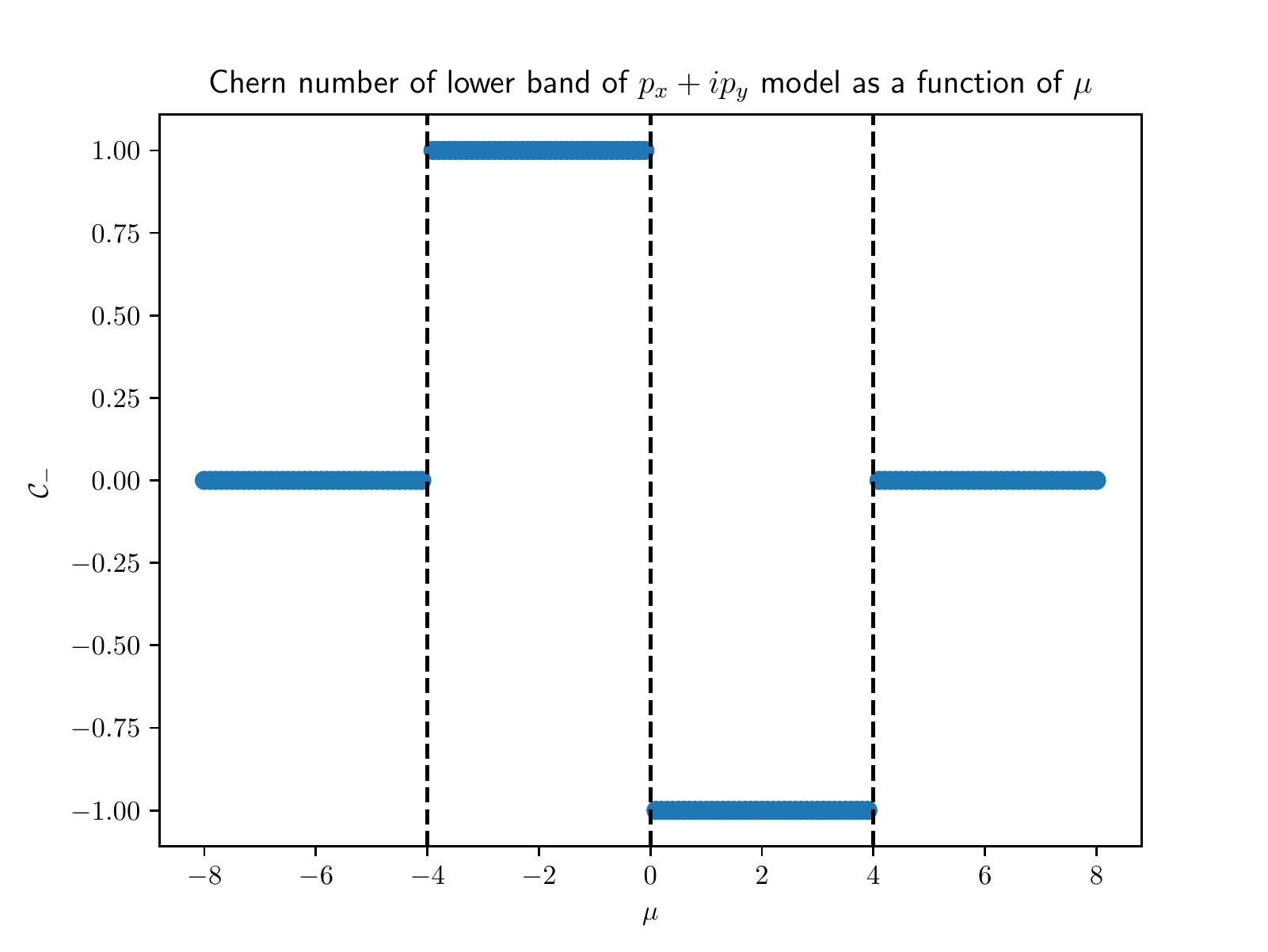}
\caption{Numerically computed Chern number of the lower band of the $p_x + i p_y$ model. The parameters $t = 1$, $\Delta = 2$ are held fixed, while $\mu$ is varied. The Chern number transitions at $\mu = - 4$, $\mu = 0$, and $\mu = 4$, in agreement with the calculations of Section \ref{sec:mod_square}.}
\label{fig:pxpy_phase}
\end{figure}

For $\Delta' \neq 0$ and $\mu' \notin \{-2,2\}$, and for $\Delta' = 0$ and $|\mu'| > 2$, the system has a spectral gap and the Chern numbers $\mathcal{C}_\pm$ of the bands $E_\pm$ are well-defined. We define the \ul{Chern numbers} of the upper and lower bands by
\begin{equation} \label{eq:chern_num}
\begin{split}
    &\mathcal{C}_\pm = \frac{1}{2 \pi i} \inty{0}{2 \pi}{ \inty{0}{2 \pi}{ \mathcal{F}_\pm(k_1,k_2) }{k_1} }{k_2}   \\
    &\mathcal{F}_\pm(k_1,k_2) =  \de_{k_1} \ip{\chi_\pm}{\de_{k_2} \chi_\pm} - \de_{k_2} \ip{\chi_\pm}{\de_{k_1} \chi_\pm}.
\end{split}
\end{equation}
When $\Delta' = 0$, time-reversal symmetry in the form
\begin{equation}
    \overline{H(k_1,k_2)} = H(-k_1,-k_2)
\end{equation}
holds, which implies both Chern numbers are trivially zero. When $\Delta' \neq 0$, numerical computation (using \cite{2005FukuiHatsugaiSuzuki}) shows that for $0 < \mu' < 2$, $\mathcal{C}_\pm = \pm 1$, for $-2 < \mu' < 0$, $\mathcal{C}_\pm = \mp 1$. When $|\mu'| > 2$, $\mathcal{C}_\pm = 0$, see Figure \ref{fig:pxpy_phase}.

\subsection{$p_x + i p_y$ model on an arbitrary finite point set of $\field{R}^2$ and disorder} \label{sec:mod_general} 
We now introduce a generalization of the model introduced in the previous section defined on arbitrary point sets of $\field{R}^2$. In what follows we will be interested in studying the $p_x + i p_y$ model introduced in Section \ref{sec:mod_square} truncated to a finite square lattice, and how properties of this system change as we add defects and disorder.

We consider a set of $N$ points in the plane with co-ordinates $r_j = (x_j,y_j) \in \field{R}^2$ where $j \in \{1,...,N\}$. We assume the set is disjoint, so that $j \neq k \iff r_j \neq r_k$. We consider the Hilbert space $\mathcal{H} := l^2\left(\{1,...,N\};\field{C}^2\right)$, denoting elements of this space by $\psi = (\psi_j)_{j \in \{1,...,N\}}$ where $\psi_j \in \field{C}^2$. We define a Hamiltonian by
\begin{equation} \label{eq:gen_Ham}
    (H \psi)_j = H_{jj} \psi_j + \sum_{\substack{k = 1 \\ j \neq k}}^N H_{jk} \psi_k,
\end{equation}
where $H_{jj}$ and $H_{jk}$ are the $2 \times 2$ matrices defined by
\begin{equation}
\begin{split}
    &H_{jj} = - \mu_j \sigma_3    \\
    &H_{jk} = \rho(|r_j - r_k|) \left[ - t \sigma_3 - \frac{i \Delta}{2} \sigma_1 \cos( \theta(r_j,r_k) ) - \frac{ i \Delta }{2} \sigma_2 \sin(\theta(r_j,r_k)) \right].
\end{split}
\end{equation}
Here the function $\rho(\zeta)$ is a kind of cutoff function defined by
\begin{equation} \label{eq:cutoff}
    \rho(\zeta) = \begin{cases} \frac{ \sqrt{2} - \zeta }{ \sqrt{2} - 1 } & 0 \leq \zeta \leq \sqrt{2} \\ 0 & \zeta > \sqrt{2}. \end{cases}  \\
\end{equation}
The function $\theta(r_j,r_k)$ denotes the angle between the vector which points from $r_j$ to $r_k$ and the unit vector pointing along the positive $x$ axis, and $\mu_j$ denotes a real number which depends on $j$. With these definitions, the action of the Hamiltonian \eqref{eq:gen_Ham} reduces to \eqref{eq:struc} in the case when the point set has the form of a regular square lattice with sides of length $1$ and all of the $\mu_j$ are equal. 

In what follows we will be interested in the dynamics of electrons modeled by the Schr\"odinger equation
\begin{equation}
    i \de_t \psi = H \psi \quad \psi(0) = \psi_0,
\end{equation}
for $\psi(t) : [0,T] \mapsto l^2(\{1,...,N\};\mathbb{C}^2)$, where $H$ is defined by \eqref{eq:gen_Ham}. We will first study these dynamics for the case of a finite point set arranged in a square lattice such that the distance between adjacent sites is $1$ and such that every $\mu_j$ is equal. We will then be interested in how these dynamics are affected by various perturbations of the atomic positions $r_j$ and onsite potentials $\mu_j$. We will first study localized, deterministic, perturbations of these parameters, and then random disorder modeled by random variations of these parameters. Note that when we perturb the atomic site positions $r_j$, the topology of the network (the pairs of sites between which hopping occurs) may change because of the distance dependence of the function $\rho$.

The specific models of disorder we will study are as follows. We suppose that the onsite potentials $\mu_j$ are drawn from a uniform distribution centered at $\mu$, with width $\sigma_\mu$. We call this kind of disorder \ul{potential disorder}. We then suppose that the site positions $r_j$ are determined by first drawing an angle $\Phi$ from the uniform distribution on the interval $[0,2 \pi]$, and then a radius $R$ from the uniform distribution on the interval $[0,\sigma_r]$. We then take
\begin{equation}
    r_j = r_{j,0} + R ( \cos(\Phi) , \sin(\Phi) ),
\end{equation}
where $r_{j,0}$ denotes the position of the $j$th site in a perfect square lattice with sides of length $1$. We call this kind of disorder \ul{position disorder}. Note that when $\sigma_\mu = \sigma_r = 0$ we recover a finite square lattice structure without disorder.

Note that in all cases other than an infinite periodic lattice, the Chern numbers in momentum space \eqref{eq:chern_num} are not well-defined. 
In the next section we will describe the spectral localizer index, a local topological index which can be used to probe properties of this system.

\section{The spectral localizer spectrum and index} \label{sec:index}

In this section, for the reader's convenience, we give a self-contained introduction to the spectral localizer and its associated index (see also \cite{2015Loring,2019DeBonisLoringSverdlov}). We introduce these concepts first without reference to the physical system we intend to study, before explaining the connection in Section \ref{sec:BB_correspondence}. The spectral localizer spectrum was originally introduced by Kisil \cite{1996Kisil}, who referred to it as the Clifford spectrum.

\subsection{Spectral localizer spectrum} \label{sec:Cliff} It is well-known that $M$ Hermitian matrices $A_j \in \field{C}^{N \times N}$, $j \in \{1,...,M\}$, which pair-wise commute:
\begin{equation} \label{eq:pw_comm}
	[A_i,A_j] = 0 \text{ for each } 1 \leq i \leq M, \; 1 \leq j \leq M,
\end{equation}
can be simultaneously diagonalized, i.e. there exists a basis of simultaneous eigenvectors $\left\{v_k\right\}_{k \in \{1,...,N\}}$ which satisfy 
\begin{equation}
    A_j v_k = \lambda_{j,k} v_k \quad j \in \{1,...,M\}, k \in \{1,...,N\}
\end{equation}
for some real numbers $\lambda_{j,k}$.

Suppose we are given Hermitian matrices $A_j \in \field{C}^{N \times N}, j \in \{1,...,M\}$, which, instead of pair-wise commuting \eqref{eq:pw_comm}, \ul{pair-wise $\delta$-commute} for some $\delta > 0$:
\begin{equation} \label{eq:alm_comm}
	\| [A_i,A_j] \| \leq \delta \text{ for each } 1 \leq i \leq m, 1 \leq j \leq m,
\end{equation}
where $\| A \|$ denotes the usual operator norm of a matrix. In this case, there is no hope of finding a basis of simultaneous eigenvectors of all of the $A_j$. We ask instead if we can find \ul{$\epsilon$-approximate eigenvectors} $w$ which satisfy:
\begin{equation} \label{eq:app_evec}
    \| A_j w - \mu_{j} w \| \leq \epsilon(\delta) \quad j \in \{1,...,M\},
\end{equation}
for some real numbers $\mu_{j}$, such that $\epsilon(\delta) \rightarrow 0$ as $\delta \rightarrow 0$.

We focus on the case $M = 3$ since this is the case which is relevant for the two-dimensional model we consider in this work. Following Loring \cite{2015Loring}, we consider the family of $2 N \times 2 N$ Hermitian matrices known as the \ul{spectral localizer}: 
\begin{equation} \label{eq:B_matrix}
	L_{\lambda}(A_1,A_2,A_3) := \begin{pmatrix} (A_3 - \lambda_3) & (A_1 - \lambda_1) - i (A_2 - \lambda_2) \\ (A_1 - \lambda_1) + i (A_2 - \lambda_2) & - (A_3 - \lambda_3) \end{pmatrix},
\end{equation}
parameterized by the vector of real numbers $\lambda = (\lambda_1, \lambda_2, \lambda_3) \in \field{R}^3$. We claim that whenever $L_{\lambda}(A_1,A_2,A_3)$ has an eigenvalue equal to zero, we can find an $\epsilon$-approximate simultaneous eigenvector of the matrices $A_i$ with associated eigenvalues $\lambda_i$ such that $\epsilon(\delta) \rightarrow 0$ as $\delta \rightarrow 0$. We refer to the set of $(\lambda_1,\lambda_2,\lambda_3) \in \mathbb{R}^3$ such that $L_{\lambda}(A_1,A_2,A_3)$ has an eigenvalue equal to zero as the \ul{spectral localizer spectrum} of the matrices $\{ A_i \}_{1 \leq i \leq 3}$. In what follows we will often abbreviate terminology and simply refer to the \ul{localizer} and \ul{localizer spectrum}.

In order to avoid writing out large matrices in detail we at this point recognize that the localizer can be written compactly using the tensor product and Pauli matrices as
\begin{equation}
	L_{\lambda}(A_1,A_2,A_3) = \sigma_1 \otimes (A_1 - \lambda_1) + \sigma_2 \otimes (A_2 - \lambda_2) + \sigma_3 \otimes (A_3 - \lambda_3).
\end{equation}

Suppose that for some value of $\lambda_1, \lambda_2$, and $\lambda_3$ such a zero eigenvalue exists: 
\begin{equation} \label{eq:B_eq}
	L_\lambda(A_1,A_2,A_3) \begin{pmatrix} v \\ w \end{pmatrix} = 0,
\end{equation}
for some $v$, $w \in \mathbb{C}^n$ where at least one of $v$ and $w$ is $\neq 0$. We first apply the matrix $L_\lambda(A_1,A_2,A_3)$ again to both sides of this equation to see that: 
\begin{equation} \label{eq:B_sq_eq}
	L_\lambda(A_1,A_2,A_3)^2 \begin{pmatrix} v \\ w \end{pmatrix} = 0.
\end{equation}
Because of the special form of $L_\lambda(A_1,A_2,A_3)$, we have that 
\begin{equation}
\begin{split}
    L_\lambda(A_1,A_2,A_3)^2 = &I \otimes \left[ (A_1 - \lambda_1)^2 + (A_2 - \lambda_2)^2 + (A_3 - \lambda_3)^2 \right]   \\
    &
    + i \sigma_1 \otimes [A_2,A_3] + i \sigma_2 \otimes [A_3,A_1] + i \sigma_3 \otimes [A_1,A_2],
\end{split}
\end{equation}
where $I$ is the $2 \times 2$ identity matrix. We at this point assume (without loss of generality since the case where $|w| \geq |v|$ is similar) that $|v| \geq |w|$. Observe that the first entry of the vector identity \eqref{eq:B_sq_eq} can be written: 
\begin{equation}
\begin{split}
	&\left( (A_1 - \lambda_1)^2 + (A_2 - \lambda_2)^2 + (A_3 - \lambda_3)^2 \right) v  \\
        &\phantom{blablablablablablablabla}= - i [A_1,A_2] v - \left( [A_3,A_1] + i [A_2,A_3] \right) w.   \\
\end{split}
\end{equation}
Dividing both sides of this equation by $|v|$ yields:
\begin{equation}
\begin{split}
	&\left( (A_1 - \lambda_1)^2 + (A_2 - \lambda_2)^2 + (A_3 - \lambda_3)^2 \right) \hat{v}  \\
        &\phantom{blablablablablablablabla}= - i [A_1,A_2] \hat{v} - \left( [A_3,A_1] + i [A_2,A_3] \right) \frac{w}{|v|},
\end{split}
\end{equation}
where $\hat{v} := \frac{v}{|v|}$. Taking the inner product of both sides of this equation gives:
\begin{equation}
\begin{split}
	&\ip{ \hat{v} }{ (A_1 - \lambda_1)^2 \hat{v} } + \ip{ \hat{v} }{ (A_2 - \lambda_2)^2 \hat{v} } + \ip{\hat{v}}{ (A_3 - \lambda_3)^2 \hat{v} } \\
	&= - i \ip{\hat{v}}{ [A_1,A_2] \hat{v}} - \ip{\hat{v}}{ [A_3,A_1] \frac{w}{|v|} } - i \ip{\hat{v}}{ [A_2,A_3] \frac{w}{|v|} }.
\end{split}
\end{equation}
Using the facts that: (1) $A_i - \lambda_i$ is Hermitian (2) $\ip{v}{v} = |v|^2$ for any vector $v \in \mathbb{C}^n$, we find that
\begin{equation}
\begin{split}
	&|(A_1 - \lambda_1) \hat{v}|^2 + |(A_2 - \lambda_2) \hat{v}|^2 + |(A_3 - \lambda_3) \hat{v}|^2 	\\
	&= - i \ip{\hat{v}}{ [A_1,A_2] \hat{v}} - \ip{\hat{v}}{ [A_3,A_1] \frac{w}{|v|} } - i \ip{\hat{v}}{ [A_2,A_3] \frac{w}{|v|} }.
\end{split}
\end{equation}
We now bound the right-hand side using the Cauchy-Schwarz inequality ($|\ip{v}{w}| \leq |v| |w|$ for any $v, w \in \mathbb{C}^n$) and the bound \eqref{eq:alm_comm} on the operator norm of the commutators:
\begin{equation} \label{eq:key_est}
\begin{split}
    &\left| (A_1 - \lambda_1) \hat{v} \right|^2 + \left| (A_2 - \lambda_2) \hat{v} \right|^2 + \left| (A_3 - \lambda_3) \hat{v} \right|^2  \\
    &\phantom{blablablablablablablabla}\leq \| [A_1,A_2] \| + \| [A_3,A_1] \| + \| [A_2,A_3] \| \leq 3 \delta.
\end{split}
\end{equation}
Since the left-hand side is a sum of positive numbers we conclude that: 
\begin{equation} \label{eq:bound}
	\left| A_i \hat{v} - \lambda_i \hat{v} \right| \leq \sqrt{3} \delta^{\frac{1}{2}} \text{ for each $1 \leq i \leq m$},
\end{equation}
where $\hat{v} \neq 0$, i.e. that $\hat{v}$ is an $\epsilon$-approximate simultaneous eigenvector of the $A_i$s with associated eigenvalues $\lambda_i$ with $\epsilon(\delta) = \sqrt{3} \delta^{\frac{1}{2}}$. 

\subsection{The spectral localizer index: a guide to the location of localizer spectrum} \label{sec:clif_ind}

We can associate an index to any point in parameter space $\lambda = (\lambda_1,\lambda_2,\lambda_3) \in \mathbb{R}^3$ not in the localizer spectrum (i.e. where $L_{\lambda}(A_1,A_2,A_3)$ does not have a zero eigenvalue) as follows. We define the \ul{spectral localizer index} of such a point to be:
\begin{equation} \label{eq:index}
	\text{ind}(\lambda_1,\lambda_2,\lambda_3) := \frac{1}{2} \text{sig\,} L_{\lambda}(A_1,A_2,A_3),
\end{equation}
where the matrix signature is defined by
\begin{equation}
\begin{split}
    &\text{sig\,} L_\lambda(A_1,A_2,A_3)  \\
    &\phantom{bla}= \left\{ \text{number of $+$ve eigenvalues of $L_\lambda$} \right\} - \left\{ \text{number of $-$ve eigenvalues of $L_\lambda$} \right\}.
\end{split}
\end{equation}
Note that whenever the index is well-defined it must be equal to an integer. Note also that the index is more properly associated to the matrix triple $(A_1 - \lambda_1,A_2 - \lambda_2,A_3 - \lambda_3)$, but when the matrices $\{A_i\}_{1 \leq i \leq 3}$ are understood it is simpler to think of the index being associated to the point $\lambda = (\lambda_1,\lambda_2,\lambda_3) \in \mathbb{R}^3$.

The localizer index can be used to detect the approximate locations of zero eigenvalues of $L_\lambda(A_1,A_2,A_3)$ as follows. We first note that the eigenvalues of a Hermitian matrix which depends on parameters vary continuously as the parameters are varied \cite{Kato}. Now suppose that we can identify two locations in parameter space $(\lambda_1,\lambda_2,\lambda_3) \in \mathbb{R}^3$ and $(\lambda_1',\lambda_2',\lambda_3') \in \mathbb{R}^3$ such that $\text{ind}(\lambda_1,\lambda_2,\lambda_3) = 1$ but $\text{ind}(\lambda_1',\lambda_2',\lambda_3')= 0$, and consider any continuous path in parameter space which starts at $(\lambda_1,\lambda_2,\lambda_3)$ and ends at $(\lambda_1',\lambda_2',\lambda_3')$. For example, we can take the straight line path:
\begin{equation}
\begin{split}
	&\left(\lambda_1(t),\lambda_2(t),\lambda_3(t)\right) := \\ 
        &\phantom{bla}\left(t \lambda_1 + (1 - t) \lambda_1',t \lambda_2 + (1 - t) \lambda_2',t \lambda_3 + (1 - t) \lambda_3' \right) \quad t \in [0,1].
\end{split}
\end{equation}
Since the path is assumed continuous, the eigenvalues of $L_\lambda(A_1,A_2,A_3)$ must vary continuously along the path, and hence the index \eqref{eq:index} must also vary continuously wherever it is defined along the path. Since any function equal to an integer which is continuous must actually be constant, we can conclude immediately that \emph{there must be a point along the path where the index is not well-defined}, i.e. there must be a point along the path which is in the localizer spectrum (where the matrix $L_\lambda(A_1,A_2,A_3)$ has a zero eigenvalue). In fact, we can make the stronger statement that at some point it must be that an eigenvalue actually transitions along the path from being a positive to a negative eigenvalue (or vice versa). If this were not the case then the index \eqref{eq:index} would not change between the ends of the path. 

The values of the index \eqref{eq:index} as $\lambda_1, \lambda_2, \lambda_3$ is varied can therefore be a guide to the location of points in the localizer spectrum which, as we have seen in the previous section, correspond to ``eigenvalues'' of $\epsilon$-approximate simultaneous eigenvectors of the matrices $\{ A_i \}_{1 \leq i \leq 3}$ (in the sense of \eqref{eq:app_evec}) where $\epsilon(\delta) \rightarrow 0$ as $\delta \rightarrow 0$.

\subsection{The localizer pseudo-spectrum: a computable alternative to the localizer spectrum}

Recall that the localizer spectrum is defined as the set of $\lambda = (\lambda_1,\lambda_2,\lambda_3) \in \mathbb{R}^3$ such that the matrix $L_\lambda(A_1,A_2,A_3)$ \eqref{eq:B_matrix} has a zero eigenvalue. It is common when computing eigenvalues of matrices numerically for the computed eigenvalues to deviate slightly from their true values because of round-off error which is inevitable when doing floating point arithmetic \cite{TrefethenBau}. It is therefore better in practice to compute, instead of the localizer spectrum, the \ul{localizer $\mu$-pseudo-spectrum} defined by: 
\begin{equation}
	\{ \lambda = (\lambda_1, \lambda_2, \lambda_3) \in \mathbb{R}^3 : L_\lambda(A_1,A_2,A_3) \text{ has an eigenvalue with norm $\leq \mu$} \}.
\end{equation}
Here $\mu \geq 0$ is assumed small, with its precise value to be chosen depending on the application. Note that when $\mu = 0$ we recover the definition of the localizer spectrum and that the localizer spectrum is clearly a subset of the localizer $\mu$-pseudo-spectrum for any $\mu \leq 0$. An essentially identical argument to that given in Section \ref{sec:clif_ind} then implies that whenever $(\lambda_1, \lambda_2, \lambda_3) \in \mathbb{R}^3$ is in the localizer $\mu$-pseudo-spectrum, the matrices $\{ A_i \}_{1 \leq i \leq 3}$ have an $\epsilon$-approximate simultaneous eigenvector with:
\begin{equation}
	\epsilon(\delta,\mu) = \left( \mu^2 + 3 \delta \right)^{\frac{1}{2}},
\end{equation}
which clearly $\rightarrow 0$ as $\mu \rightarrow 0$ and $\delta \rightarrow 0$. 

We note finally that to ensure that the localizer index is truly well defined whenever we numerically compute it we should assign a localizer index only to points not in the localizer $\mu$-\emph{pseudo}-spectrum for $\mu > 0$. Note that this doesn't change the conclusion of the argument given at the end of Section \ref{sec:clif_ind}. If the localizer index differs at the ends of a continuous path in parameter space, at least one point along the path must lie in the localizer \emph{spectrum}, since for the index to change it must be that an eigenvalue transitions from being positive to negative (or vice versa). Since the eigenvalues of $L_\lambda(A_1,A_2,A_3)$ must vary continuously as a function of the parameters $\lambda_i$, we conclude that there must be a point along the path where $L_\lambda$ has an eigenvalue which is \emph{precisely} zero, i.e. a point in the localizer spectrum. On the other hand, we are limited in our ability to actually compute the precise location of this point accurately: doing so would imply the ability to numerically compute eigenvalues with arbitrary precision which is impossible. 

\subsection{The localizer gap: a measure of the ``strength'' of the localizer index} \label{sec:strength}

It can be important to know, given a fixed $\lambda^1 = (\lambda_1^1,\lambda_2^1,\lambda_3^1)$ \emph{not} in the localizer pseudo-spectrum, whether there exists $\lambda^2 = (\lambda_1^2,\lambda_2^2,\lambda_3^3)$ \emph{nearby} to $\lambda^1$ (with respect to the Euclidean norm) such that ${\lambda^2}$ \emph{is} in the localizer pseudo-spectrum. An indicator of this is the size of the smallest magnitude eigenvalue of $L_{\lambda^1}(A_1,A_2,A_3)$, which we will refer to as the \ul{localizer gap}. Note that by assumption on $\lambda^1$ this quantity is non-zero. 

The reason the smallest magnitude eigenvalue of the localizer is important is as follows. 
As $\lambda \rightarrow {\lambda^2}$, the smallest magnitude eigenvalue of $L_\lambda(A_1,A_2,A_3)$ must become small. However, since the eigenvalues of $L_{\lambda}(A_1,A_2,A_3)$ vary \emph{continuously} as functions of $\lambda$, the smallest magnitude eigenvalue cannot become arbitrarily small arbitrarily fast. Hence if the size of the smallest magnitude eigenvalue is large at $\lambda^1$, it indicates there cannot be any ${\lambda^2}$ in the localizer spectrum very close to $\lambda^1$. 

This argument can be made quantitative; see Section 7 of \cite{2015Loring}. The localizer gap can also be used to prove a quantitative version of the argument given in Section \ref{sec:clif_ind} that between points with differing indices there must be localizer spectrum. Given two points with differing indices, if the size of the localizer gap at each point is large, the location of the localizer spectrum between the two points will be very tightly constrained; see Theorem 7.5 of \cite{2015Loring}.

We find that the localizer gap conveys important information when we apply the theory of the localizer index to the $p_x + i p_y$ model below; see Section \ref{sec:strength_2}.

\subsection{Tuning the spectral localizer} \label{sec:tuning}

We have seen that whenever $\lambda$ is in the localizer spectrum we can find an $\epsilon$-approximate simultaneous eigenvector $\hat{v}$ of the matrices $A_i$. It is possible to tune the localizer so that the $\hat{v}$ found in this way is arbitrarily close to an exact eigenvector of any one of the $A_i$s, at the cost of weakening the bound \eqref{eq:bound} for the other two $A_i$s.

Suppose we want to force $\hat{v}$ to be an almost exact eigenvector of $A_3$. The idea is to replace, in the definition of $L_\lambda(A_1,A_2,A_3)$, the matrices $A_1 - \lambda_1$ and $A_2 - \lambda_2$ by $\kappa (A_1 - \lambda_1)$ and $\kappa (A_2 - \lambda_2)$. The key estimate \eqref{eq:key_est} then becomes
\begin{equation}
\begin{split}
    &\kappa^2 | (A_1 - \lambda_1) \hat{v} |^2 + \kappa^2 | (A_2 - \lambda_2) \hat{v} |^2 + | (A_3 - \lambda_3) \hat{v} |^2      \\
    &\phantom{blablablablabla} \leq \mu^2 + \kappa^2 \| [A_1,A_2] \| + \kappa \| [A_3,A_1] \| + \kappa \| [A_2,A_3] \|,
\end{split}
\end{equation}
from which we can estimate
\begin{equation} \label{eq:good}
    | (A_3 - \lambda_3) \hat{v} | \leq \left( \mu^2 + \left( \kappa^2 + 2 \kappa \right) \delta \right)^{\frac{1}{2}},
\end{equation}
which can be made arbitrarily small for fixed $\delta$ by taking $\mu$ and $\kappa$ sufficiently small. When we do this we pay a price in that the estimates on $(A_1 - \lambda_1) \hat{v}$ and $(A_2 - \lambda_2) \hat{v}$ become weaker as $\kappa$ is made smaller:
\begin{equation} \label{eq:bad}
    | (A_i - \lambda_i) \hat{v} | \leq \left( \left(\frac{\mu}{\kappa}\right)^2 + \left( 1 + \frac{2}{\kappa} \right) \delta \right)^{\frac{1}{2}} \quad i = 1,2.
\end{equation}
That the bound \eqref{eq:good} can only be strengthened at the cost of weakening the bound \eqref{eq:bad} can be seen as a manifestation of the uncertainty principle. We find that tunability of the localizer is important for making the localizer predictive in the case of the $p_x + i p_y$ model; see Section \ref{sec:tuning_2}.

\section{Application of the localizer index to the $p_x + i p_y$ model: bulk-boundary correspondence} \label{sec:BB_correspondence}

We now describe how the results of the previous section can be applied to study the $p_x + i p_y$ model on a finite point set of $\field{R}^2$ introduced in Section \ref{sec:mod_general}. We will pay particular attention to the form of bulk-boundary correspondence which can be defined through the localizer.

Consider the matrices $X$, $Y$, and $H$, where $X$ and $Y$ are the position operators acting on the component of $\psi$ at position $r_j = (x_j,y_j)$ by
\begin{equation}
    \left( X \psi \right)_j = x_j \psi_j, \quad \left( Y \psi \right)_j = y_j \psi_j,
\end{equation}
and $H$ is the system Hamiltonian defined by \eqref{eq:gen_Ham}. Note that the position operators clearly commute
\begin{equation}
    [ X , Y ] = 0.
\end{equation}
Hence, by defining
\begin{equation}
    \delta := \max\left\{ \| [X,H] \|, \| [Y,H] \| \right\},
\end{equation}
we have that $X$, $Y$, $H$ are a triple of pair-wise $\delta$-commuting matrices in the sense of \eqref{eq:alm_comm}. Since the off-diagonal terms $H_{jk}$ in the Hamiltonian are zero whenever two sites at locations $r_j$ and $r_k$ are a distance larger than $\sqrt{2}$ apart (recall \eqref{eq:cutoff}), we have that $\delta$ cannot be too large.

\subsection{Bulk-boundary correspondence defined through the localizer}

We will refer to an approximate simultaneous eigenvector of the matrices $X$, $Y$, and $H$, with eigenvalues $\lambda_1$, $\lambda_2$, and $\lambda_3$ respectively (i.e. the $\hat{v}$ appearing in \eqref{eq:bound}), as a \ul{localizer state} with energy $\lambda_3$ at position $(\lambda_1,\lambda_2)$. The localizer defines a notion of bulk-boundary correspondence for localizer states as follows.

First, note that existence of points $\lambda = (\lambda_1,\lambda_2,\lambda_3)$ in the localizer spectrum of the spectral localizer $L_\lambda(X,Y,H)$ imply the existence of localizer states. On the other hand, note that points in the localizer spectrum are bound to occur at the boundary of regions of differing localizer index, by the argument given in Section \ref{sec:clif_ind}. By this argument we have the following: 
\begin{theorem*}[Bulk-boundary correspondence for localizer states]
Along any line through the space of $\lambda$ values connecting points within regions of differing localizer index, there must be at least one point $\lambda = (\lambda_1,\lambda_2,\lambda_3)$ in the localizer spectrum, and hence a localizer state with energy $\lambda_3$ at position $(\lambda_1,\lambda_2)$.
\end{theorem*}
The bulk-boundary correspondence for localizer states differs from forms of bulk-boundary correspondence which concern \emph{exact} eigenvectors of $H$. However, there is a relationship with those results. Suppose that $H$ has an exact eigenvector with eigenvalue $\lambda_3$. Then we can produce localizer states with energy $\lambda_3$ at points $(\lambda_1,\lambda_2)$ in the support of this eigenvector by multiplying the exact eigenvector of $H$ by a function which is localized in position (such as a Gaussian) centered at $(\lambda_1,\lambda_2)$. Hence the existence of exact eigenvectors of $H$, such as those whose existence is guaranteed by bulk-boundary correspondence, implies the existence of localizer states within the support of the exact eigenvector. The opposite implication is not necessarily true, however. The existence of localizer states does not necessarily imply the existence of exact eigenvectors of $H$ supported in the same location.


The localizer index with $\lambda_3$ at the Fermi level and $(\lambda_1,\lambda_2)$ in the middle of a finite structure was recently proved rigorously to agree with the Chern number of the corresponding infinite structure when the structure is sufficiently large \cite{2018LoringSchulz-Baldes}. We investigate this numerically for the (relatively small) system size we consider in this paper below; see Section \ref{sec:loc_ind_Chern} and Figure \ref{fig:loring_phase}.

\section{Detailed problem statement and outline of remainder of paper} \label{sec:prob_stat}

We are now in a position to make the rough problem statement given in Section \ref{sec:intro} precise. We will then outline the remainder of our paper where we present our results.

The first notion we make more precise is that of a wave-packet. In this work we will not form wave-packets either by ``localizing'' exact eigenvectors of $H$ by multiplying by a localized function such as a Gaussian, or by super-posing exact eigenvectors of $H$. We will instead work directly with the localizer states of $X$, $Y$, and $H$, i.e. the vector $\hat{v}$ which appears in \eqref{eq:bound}. Given our interest in the localizer index, this choice is clearly natural. In any case, we find that these approximate eigenvectors propagate similarly to wave-packets formed by ``localizing'' (i.e. multiplying by a Gaussian) exact eigenvectors.

The problem we are interested in is as follows. 
\begin{problemstatement*}
Suppose we are given a structure, described by a Hamiltonian $H$, such that the localizer index varies between regions of the structure. By the bulk-boundary correspondence for localizer states (see Section \ref{sec:BB_correspondence}), there must exist localizer states supported at the boundary between regions of differing localizer index. {What can be said about the solution $\psi(t)$, for $t \geq 0$, of the time-dependent Schr\"odinger equation}
\begin{equation} \label{eq:schro}
    i \de_t \psi = H \psi \quad \psi(0) = \psi_0,
\end{equation}
{where $\psi_0$ is such a localizer state?} 
\end{problemstatement*}

We start from the following basic hypothesis. 
\begin{hypothesis*}
The solution remains mostly localized for $t \geq 0$, while propagating along the curve of pseudo-spectrum which separates the regions of differing localizer index on either side. 
\end{hypothesis*}
This picture of the dynamics (although perhaps not in the language of the localizer index and states) is well-established in the absence of defects and disorder in the structure. Indeed it follows immediately from the bulk-boundary correspondence for periodic structures. We ambitiously hope that such a simple picture captures the dynamics \emph{even in the presence of defects and disorder}. If this simple picture breaks down, we hope that the spectral localizer can be used to gain insight into when and why.

We will test this hypothesis in the remainder of this paper by the following steps:
\begin{enumerate}
\item We compute the localizer index and pseudo-spectrum for different realizations of the $p_x + i p_y$ model, with and without disorder (Section \ref{sec:ind_comp}).
\item We compute the dynamics of the Schr\"odinger equation \eqref{eq:schro} for the same realizations, while overlaying these dynamics on a plot of the localizer index and pseudo-spectrum (Section \ref{sec:propagation}).
\end{enumerate}

Before we present our results, it is important to remark on two details of the experiments we carry out in this work.

\subsection{Localizer gap} \label{sec:strength_2} In our plots, at each point where the localizer index is well-defined, we plot also the localizer gap (recall Section \ref{sec:strength}). In the context of the $p_x + i p_y$ model, when the localizer gap is large at some $\lambda^1 = (\lambda_1^1,\lambda_2^1,\lambda_3^1)$ it indicates that there is no localizer state at any $\lambda$ nearby to $\lambda^1$. Our experiments show that localizer gap information is crucial for understanding when our basic hypothesis breaks down. 


\subsection{Tuning the spectral localizer} \label{sec:tuning_2} Recall Section \ref{sec:tuning} on ``tuning'' the localizer. We found during our numerical experiments that tuning the localizer to yield localizer states which are more tightly localized with respect to energy made the localizer index more predictive. We did this by replacing the operators $(X - \lambda_1)$ and $(Y - \lambda_2)$ by $\kappa (X - \lambda_1)$ and $\kappa (Y - \lambda_2)$ in the definition of the localizer and taking $\kappa = .5$. Specifically, we found that this choice yielded localizer states which propagated robustly while remaining quite well localized in position even in the presence of moderate disorder.

It is not yet known how to optimally tune the localizer in general; see \cite{2019Loring} for an extensive study of this.

\section{Numerical computations of the localizer index and strength} \label{sec:ind_comp}

In this section we present computations of the localizer index, localizer gap, and pseudo-spectrum of the $p_x + i p_y$ model for various choices of parameters. 

\subsection{Finite square lattice}

\begin{figure}
\begin{subfigure}[b]{\textwidth}
\centering
\includegraphics[scale=.5,draft=false]{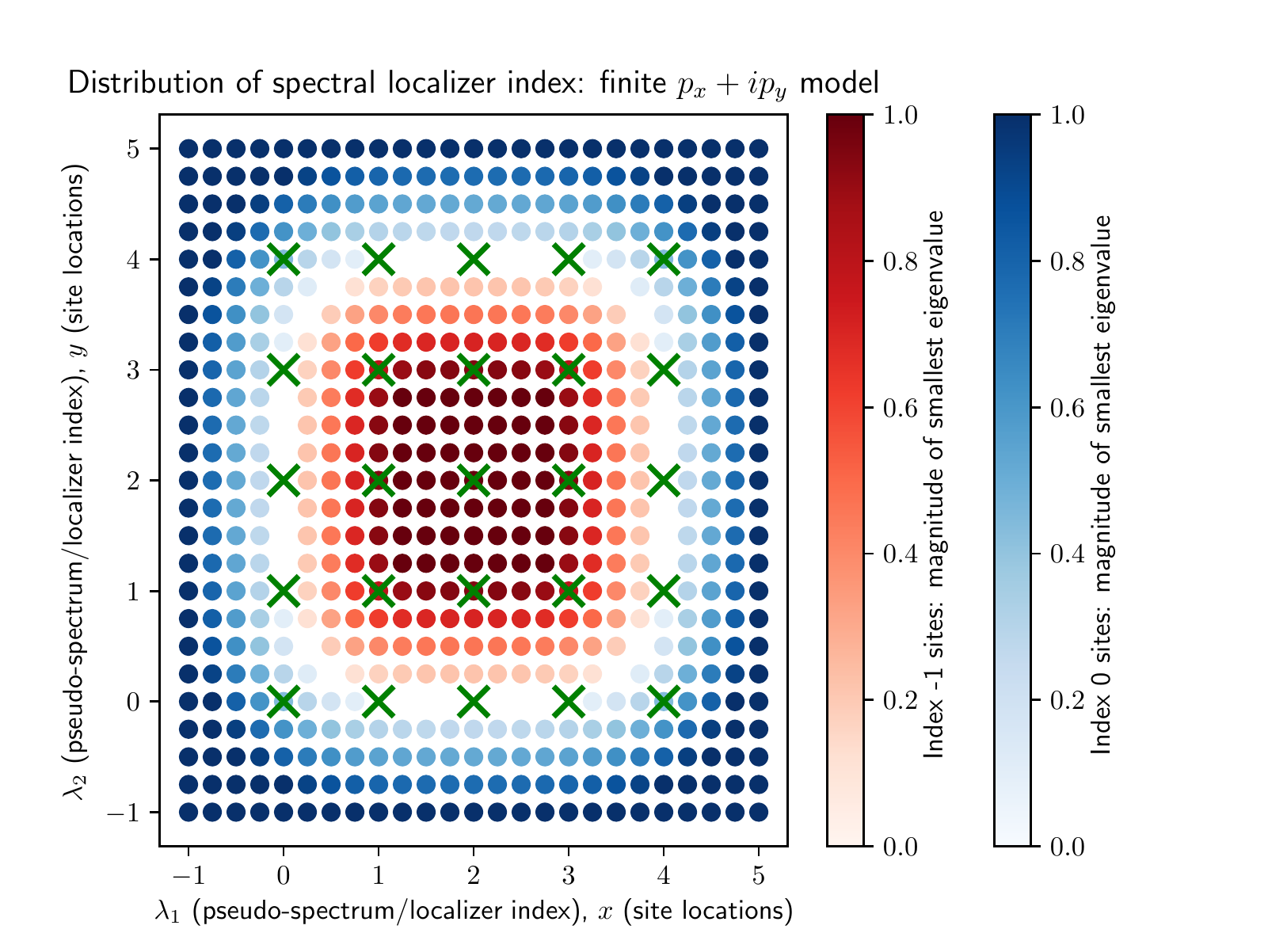}
\caption{} \label{fig:loring_example}
\end{subfigure}
\begin{subfigure}[b]{\textwidth}
\centering
\includegraphics[scale=.5,draft=false]{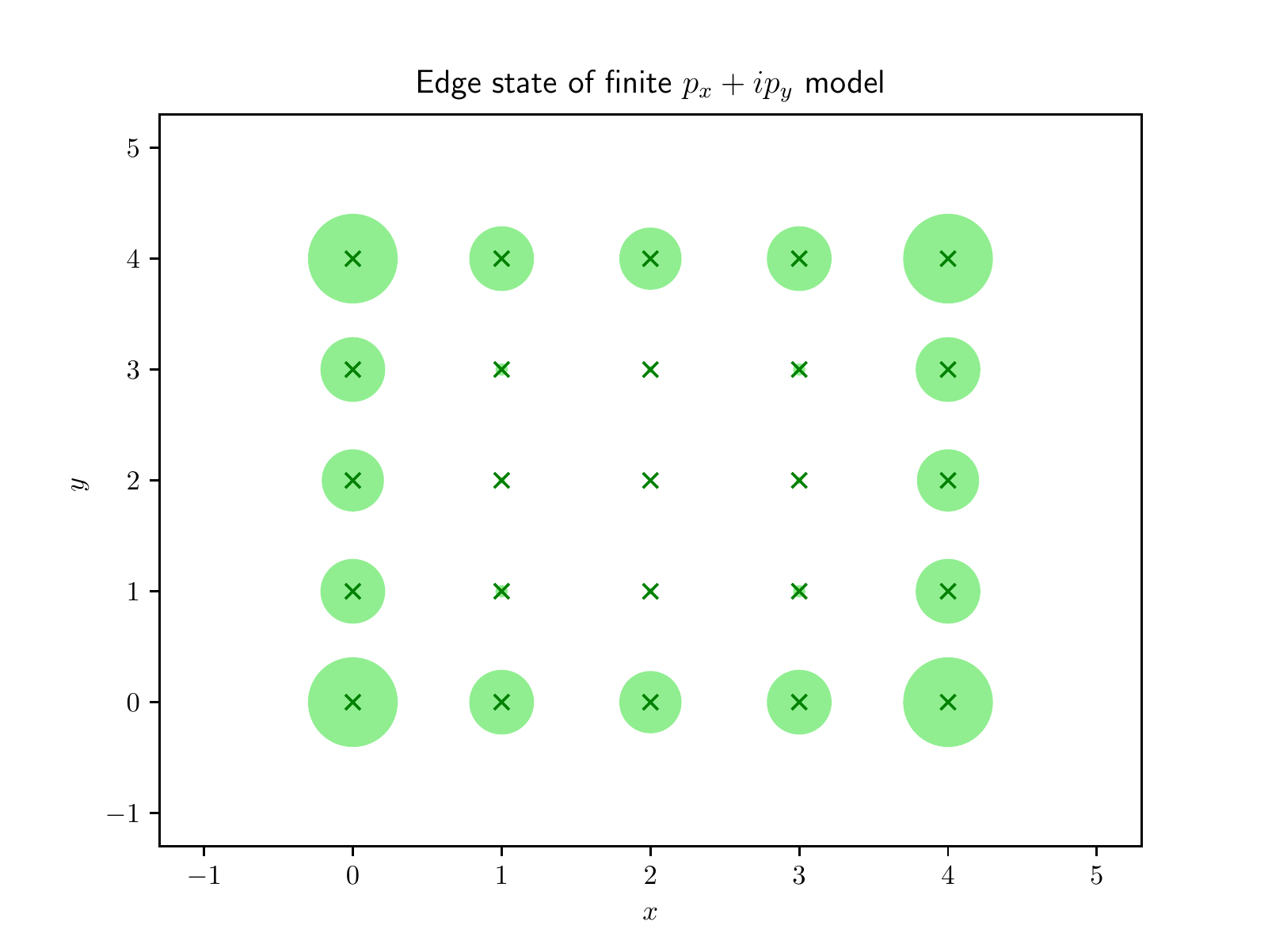}
\caption{} \label{fig:loring_example_edgestate}
\end{subfigure}
\caption{(a) Computed spectral localizer index and pseudo-spectrum at $\lambda_3 = 0$ for a finite sample modeled by the $p_x + i p_y$ model when $\mu = 2, \Delta = 2, t = 1$. Recall that the infinite periodic model has Chern number $-1$ for these values of parameters. Points $(\lambda_1,\lambda_2)$ marked white are in the localizer pseudo-spectrum, those marked blue have spectral localizer index $0$, while those marked red have spectral localizer index $-1$. The locations of atomic sites are marked by green crosses. We interpret the ring of pseudo-spectrum near to the edge of the sample as corresponding to edge states (eigenstates of $H$ localized at the edge of the sample). The saturation of the blue and red indicating the value of the local index is defined by the localizer gap: the magnitude of the smallest magnitude eigenvalue of $L_\lambda$ at that value of $(\lambda_1,\lambda_2)$. This is natural because points $(\lambda_1,\lambda_2)$ closer to points in the pseudospectrum are colored lighter than those further away. We can think of the localizer gap as indicating the ``strength'' of the local index at that point. (b) The absolute value squared of the components of an edge state of $H$ at each site, shown by the area of the green circles, with eigenvalue 0.275 (3sf).}
\end{figure}

We first consider the case of the finite square lattice with sides of length $1$, taking $t = 1, \Delta = 2$, and every $\mu_j = 2$. We find that the spectral localizer index agreed with the Chern number of the infinite periodic structure for all $(\lambda_1, \lambda_2)$ corresponding to interior points of the finite structure, and equals $0$ for $(\lambda_1, \lambda_2)$ outside the finite structure: see Figure \ref{fig:loring_example}. The pseudo-spectrum around the edge of the sample corresponds to the support of edge states: exact eigenvectors of the Hamiltonian supported around the edge of the sample. We verify the existence of edge states by diagonalizing $H$, see Figure \ref{fig:loring_example_edgestate}.

\subsection{Localizer index in center of lattice agrees with Chern number} \label{sec:loc_ind_Chern}

\begin{figure}
\includegraphics[scale=.5,draft=false]{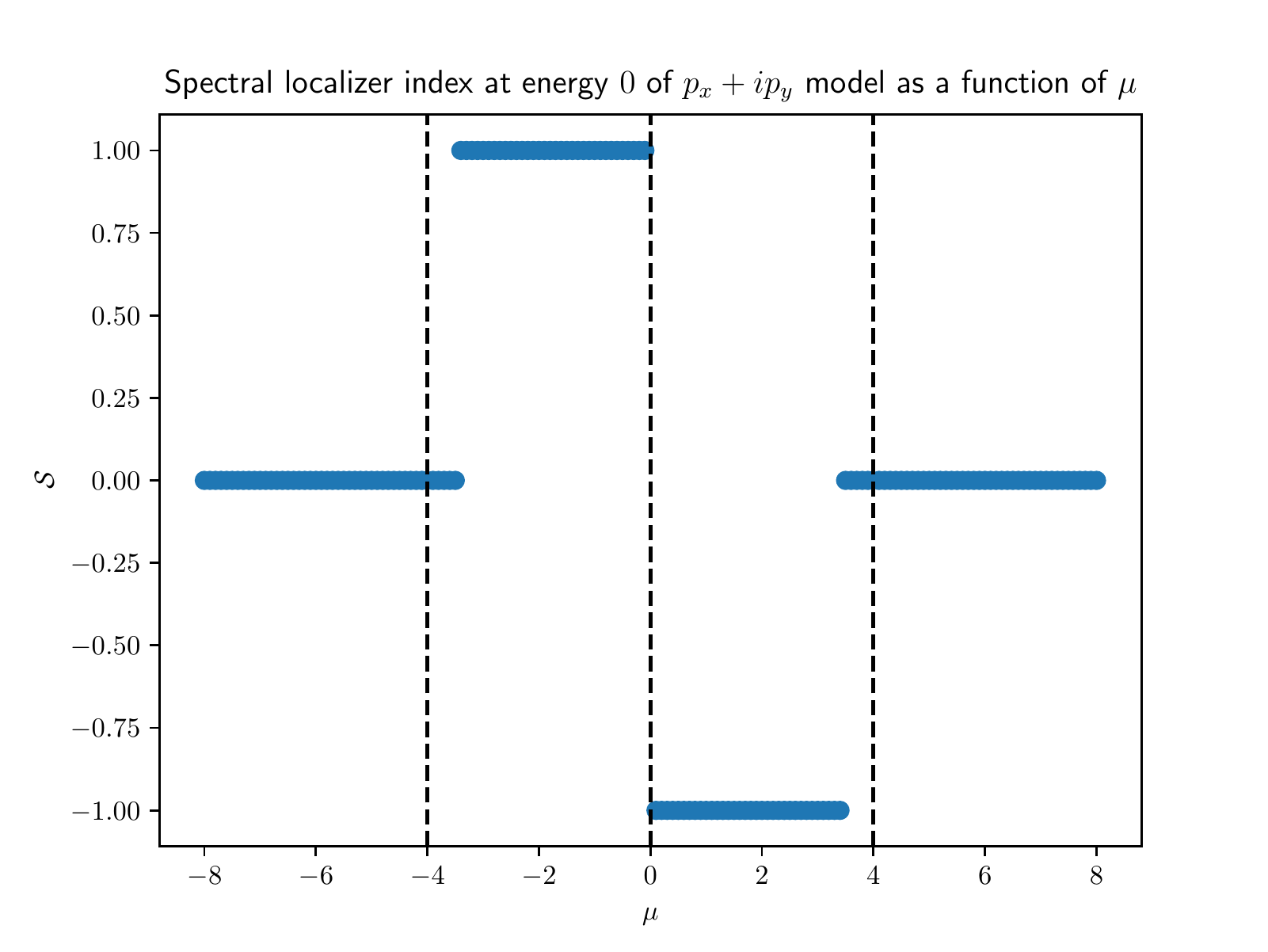}
\caption{The spectral localizer index $\mathcal{S}$ of the $p_x + i p_y$ model on a finite square lattice where $\lambda_1$ and $\lambda_2$ are chosen in the center of the sample while $\lambda_3 = 0$, i.e. in the spectral gap of $H$ when the system is extended by periodicity. The parameters $t = 1$, $\Delta = 2$ are held fixed, while $\mu$ is varied. The system size is $10 \times 10$. The spectral localizer index transitions nearby to $\mu = - 4$, $\mu = 0$, and $\mu = 4$, in agreement with our previous computations of the Chern number of the same model on an infinite periodic square lattice (Figure \ref{fig:pxpy_phase}). That the transitions do not exactly agree with Figure \ref{fig:pxpy_phase} is due to finite size effects: for sufficiently large system size the numbers are known to agree \cite{2018LoringSchulz-Baldes}.}
\label{fig:loring_phase}
\end{figure}

It is interesting to compare the value of the localizer index with $(\lambda_1,\lambda_2)$ in the center of the finite lattice and $\lambda_3 = 0$, with the Chern number of the lower band of the $p_x + i p_y$ model on an infinite square lattice. In Figure \ref{fig:loring_phase} we plot computed values of the spectral localizer index as the onsite potential $\mu$ is varied. We find that these computations agree quite well with our previous computations of the Chern number (Figure \ref{fig:pxpy_phase}). When the finite lattice size is sufficiently large, these numbers are known to exactly agree \cite{2018LoringSchulz-Baldes}.

\subsection{Perturbations of the square lattice}

We now consider two strong, deterministic perturbations of the square lattice: a perturbation where the sign of the onsite potentials is flipped in one part of the lattice, and a perturbation where some of the lattice sites are effectively removed by making their onsite potentials very large.  

\begin{figure}
\centering
\includegraphics[scale=.5,draft=false]{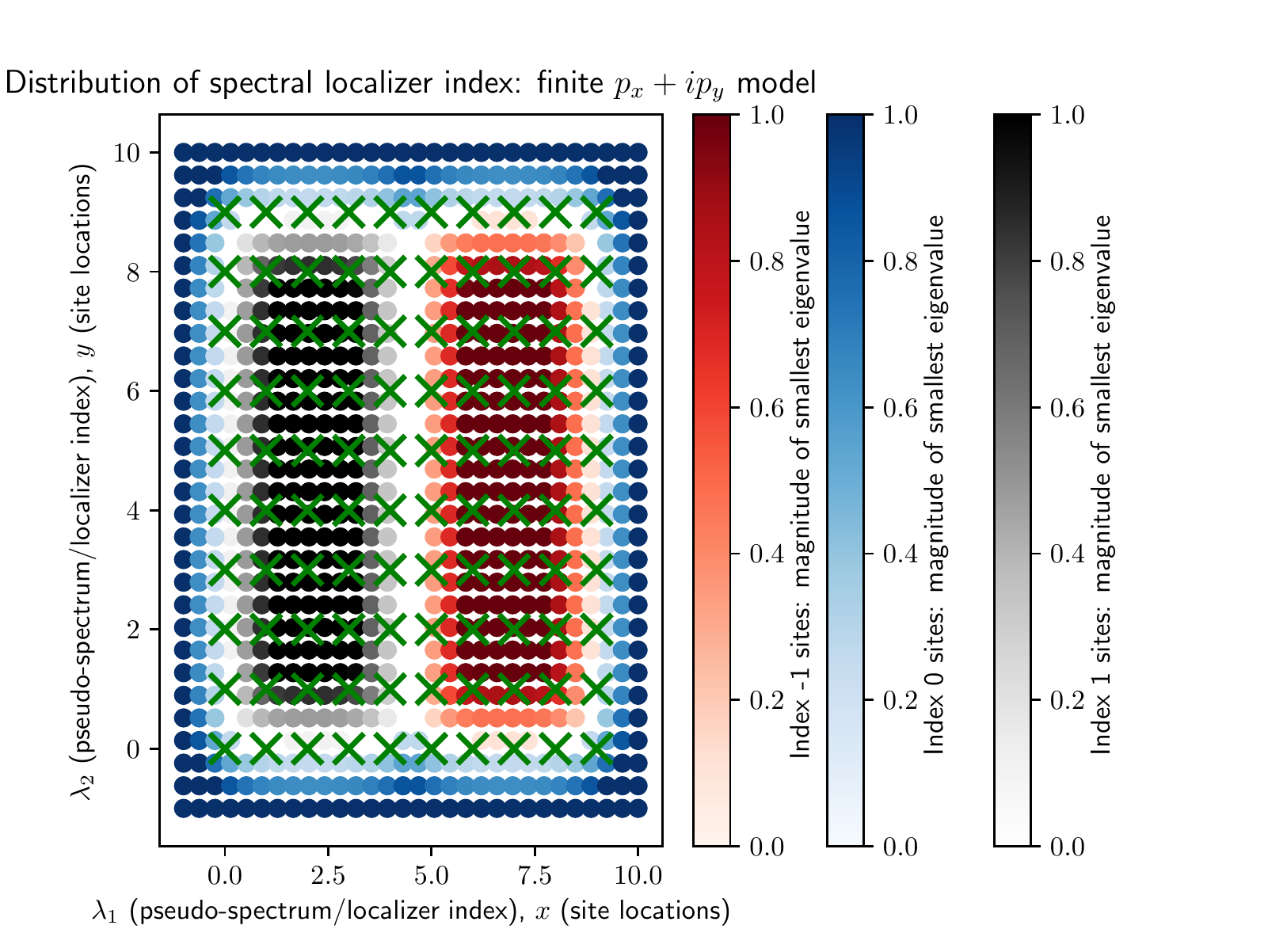}
\caption{Computed spectral localizer index and pseudo-spectrum at $\lambda_3 = 0$ for a finite sample modeled by the $p_x + i p_y$ model with $\delta = 2, t = 1$, and spatially varying $\mu$. On the left of the sample, we took $\mu = -2$. On the right, $\mu = 2$. We observe regions of different local index values: where $\mu = 2$, the index is $-1$, and where $\mu = -2$, the index is $1$. We also observe pseudospectrum along the line separating the regions of index $1$ and $-1$. We interpret this pseudospectrum as resulting from edge states supported at the interface between the topologically distinct regions.
}
\label{fig:abrupt_change}
\end{figure}

By locally changing the sign of $\mu$, we can produce structures where one part of the structure has localizer index $1$ while another has index $-1$. In this case we see pseudospectrum at the interface between the regions of differing index in addition to the pseudospectrum around the physical edge of the structure, see Figure \ref{fig:abrupt_change}. We interpret this pseudospectrum as resulting from edge states supported at the interface between the topologically distinct regions. 

\begin{figure}
\begin{subfigure}[b]{\textwidth}
\centering
\includegraphics[scale=.5,draft=false]{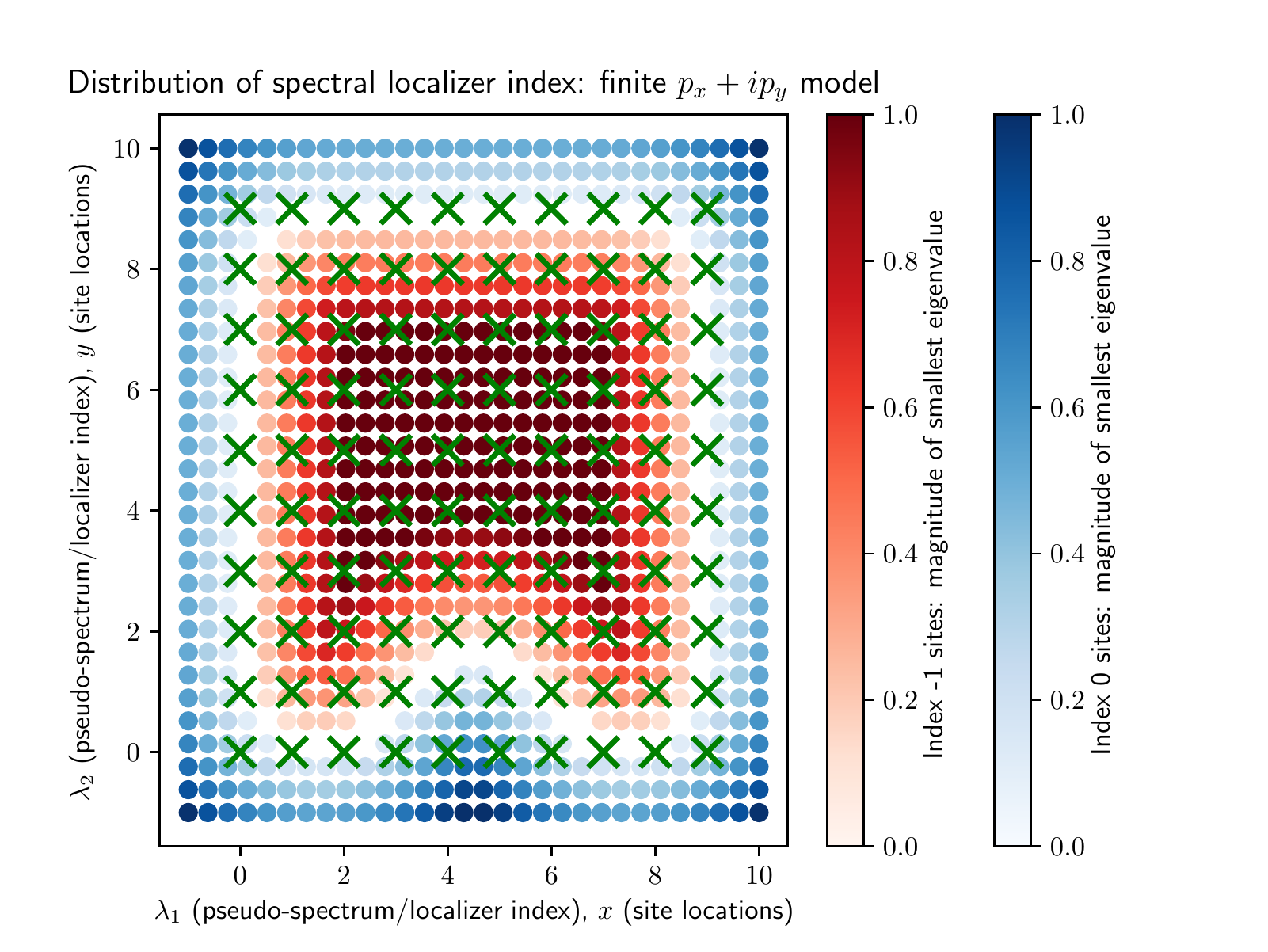}
\caption{} 
\end{subfigure}
\caption{Computed spectral localizer index and pseudo-spectrum at $\lambda_3 = 0$ for a finite sample modeled by the $p_x + i p_y$ model with $\delta = 2, t = 1$, and spatially varying $\mu$. To generate this figure, we set $\mu = 2$ everywhere other than 4 sites in the middle of the bottom edge where we set $\mu = 10^{10}$. We observe that the line of pseudospectrum around the structure moved as if these sites had actually been removed.  
}
\label{fig:defect}
\end{figure}

By making $\mu$ very large for some sites in the lattice we can effectively remove these sites from the lattice. In this case we see the pseudospectrum adjust to follow the boundary of the new structure: see Figure \ref{fig:defect}.

\subsection{Localizer index with disorder}

\begin{figure}
\centering
\includegraphics[scale=.5,draft=false]{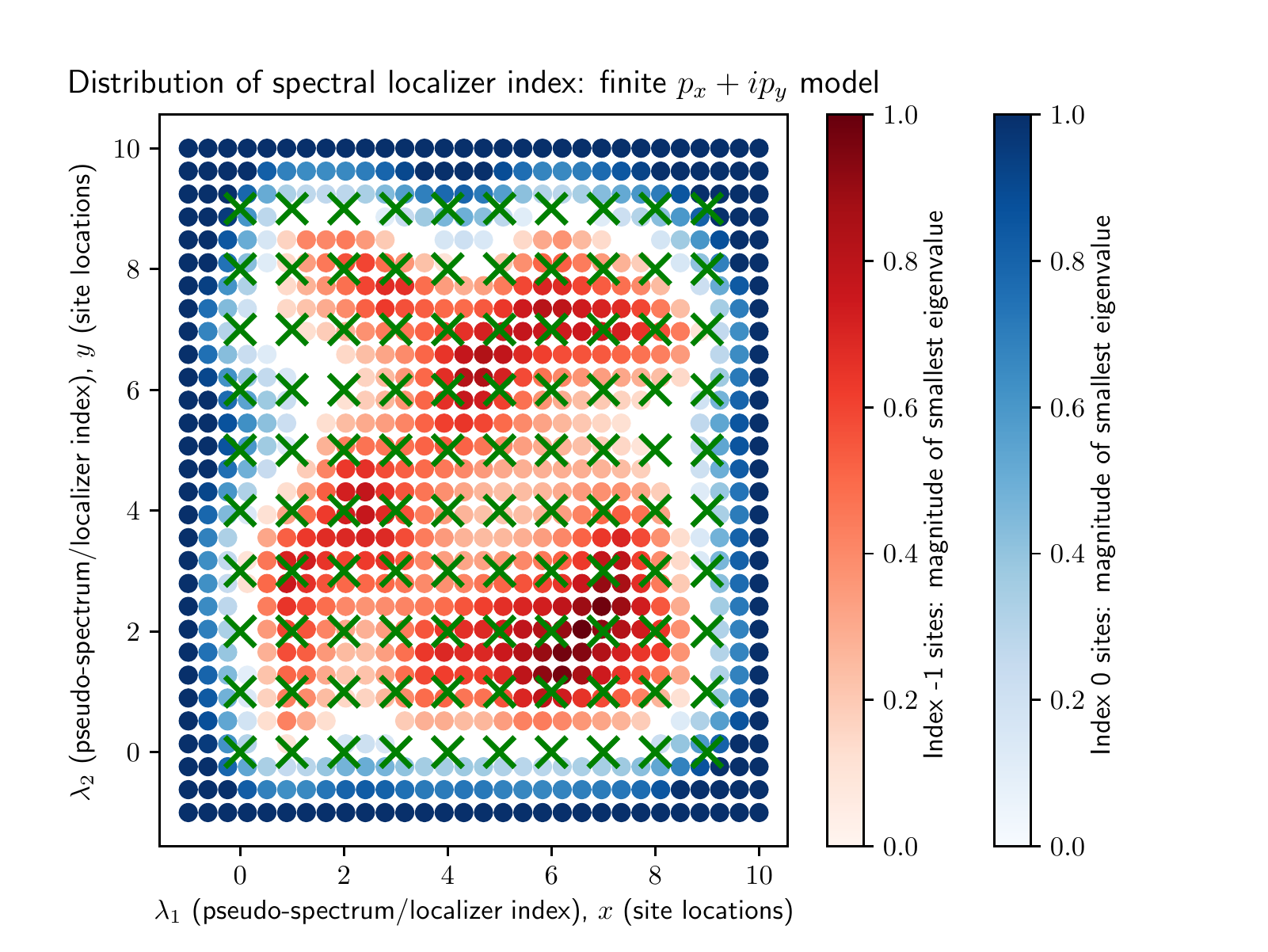}
\caption{Computed spectral localizer index and pseudo-spectrum (white) at $\lambda_3 = 0$ for a finite sample modeled by the $p_x + i p_y$ model under potential disorder. For this simulation we took $\mu = 2, \delta = 2, t = 1, \sigma_\mu = 6, \sigma_r = 0$.} 
\label{fig:pot_dis}
\end{figure}

\begin{figure}
\begin{subfigure}[b]{\textwidth}
\centering
\includegraphics[scale=.5,draft=false]{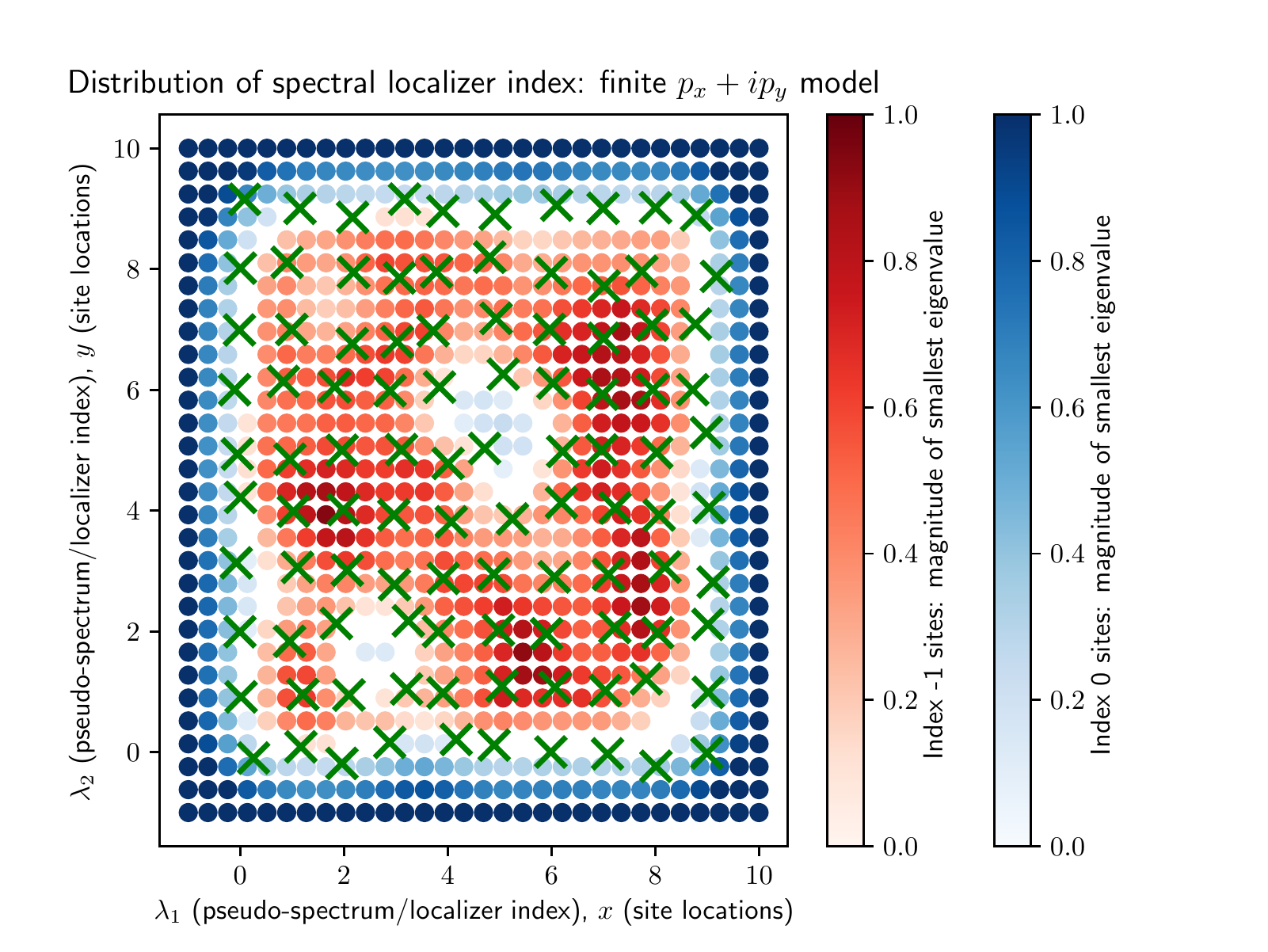}
\caption{} \label{fig:pos_dis_loring}
\end{subfigure}
\begin{subfigure}[b]{\textwidth}
\centering
\begin{subfigure}[b]{\textwidth}
\centering
\includegraphics[scale=.5,draft=false]{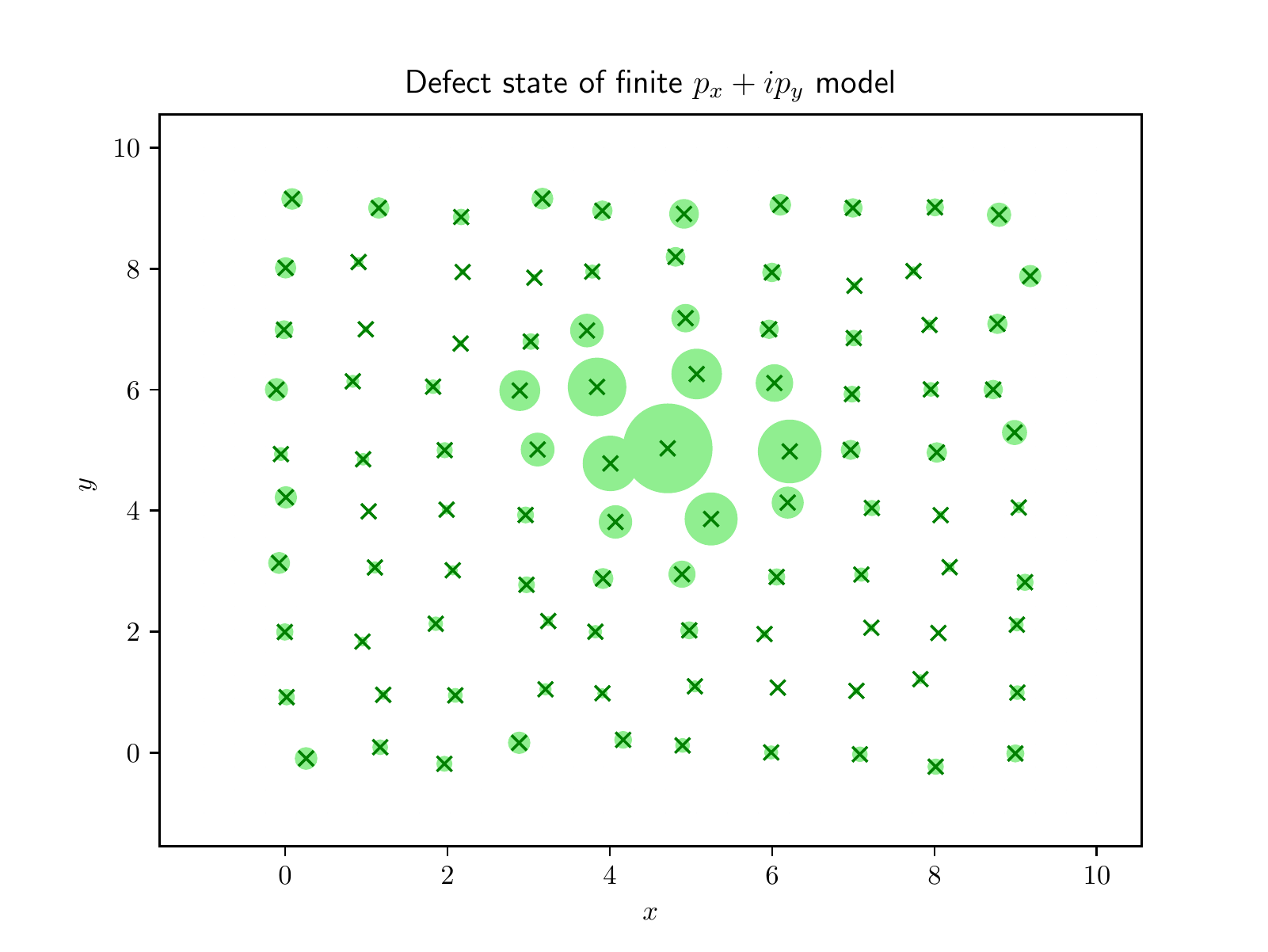}
\caption{} \label{fig:pos_dis_edgestate_3}
\end{subfigure}
\end{subfigure}
\caption{(a) Computed spectral localizer index and pseudo-spectrum (white) at $\lambda_3 = 0$ for a finite sample modeled by the $p_x + i p_y$ model under position disorder. For this simulation we took $\mu = 2, \delta = 2, t = 1, \sigma_\mu = 0, \sigma_r = .3$. Note the appearance of pseudo-spectrum and islands of index $0$ points near the center of the sample, suggesting the existence of disorder-induced defect states. (b) 
The absolute value squared of the components of a disorder-induced defect state of $H$ near the center of the sample, shown by the area of the green circles, with eigenvalue 0.563 (3sf).
} 
\label{fig:pos_dis}
\end{figure}

\begin{figure}
\centering
\includegraphics[scale=.5,draft=false]{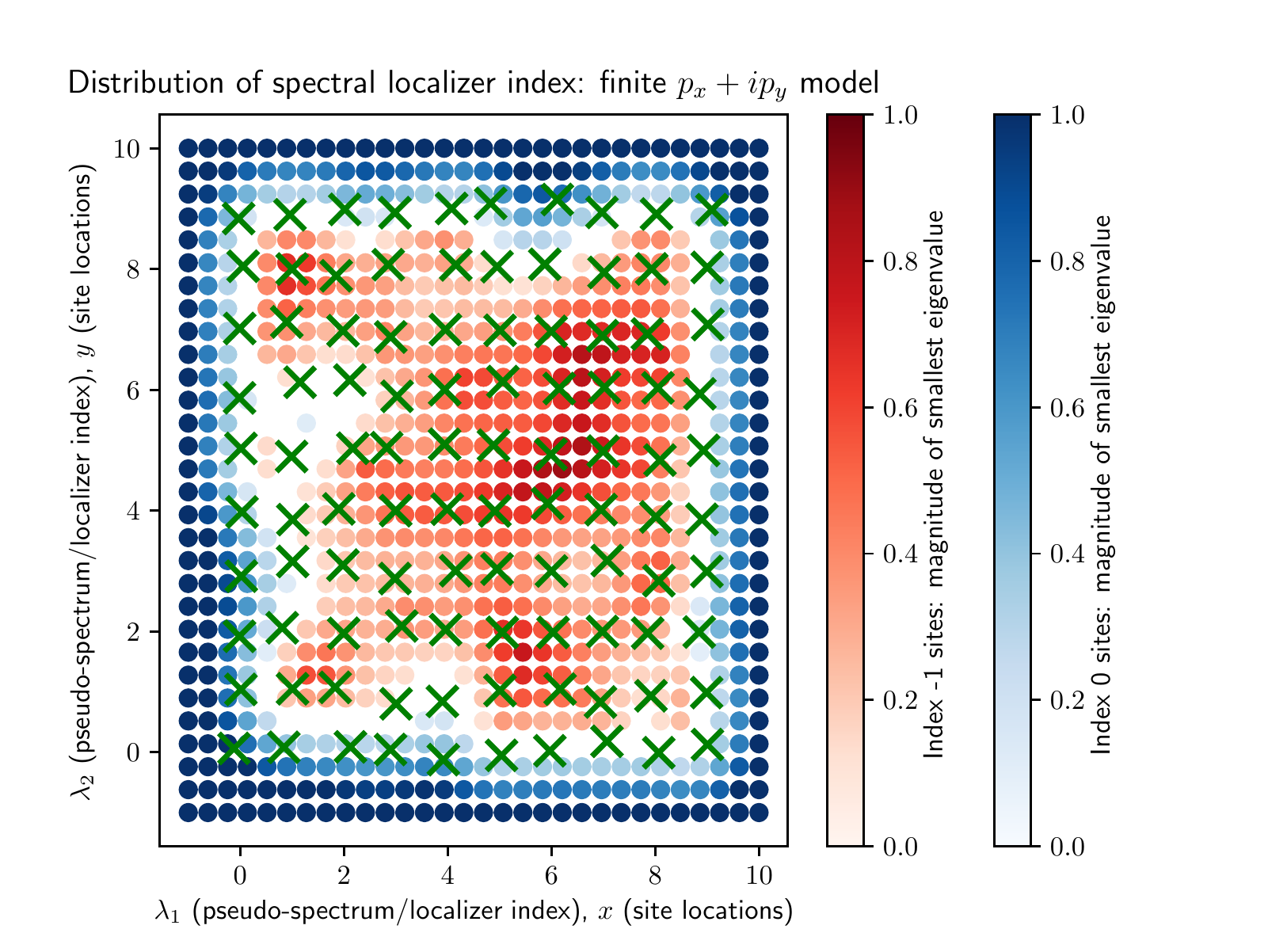}
\caption{Computed spectral localizer index and pseudo-spectrum (white) at $\lambda_3 = 0$ for a finite sample modeled by the $p_x + i p_y$ model under both potential and position disorder. For this simulation we took $\mu = 2, \delta = 2, t = 1, \sigma_\mu = 4, \sigma_r = .2$.} 
\label{fig:both_dis}
\end{figure}

\begin{figure}
\begin{subfigure}[b]{\textwidth}
\centering
\includegraphics[scale=.5,draft=false]{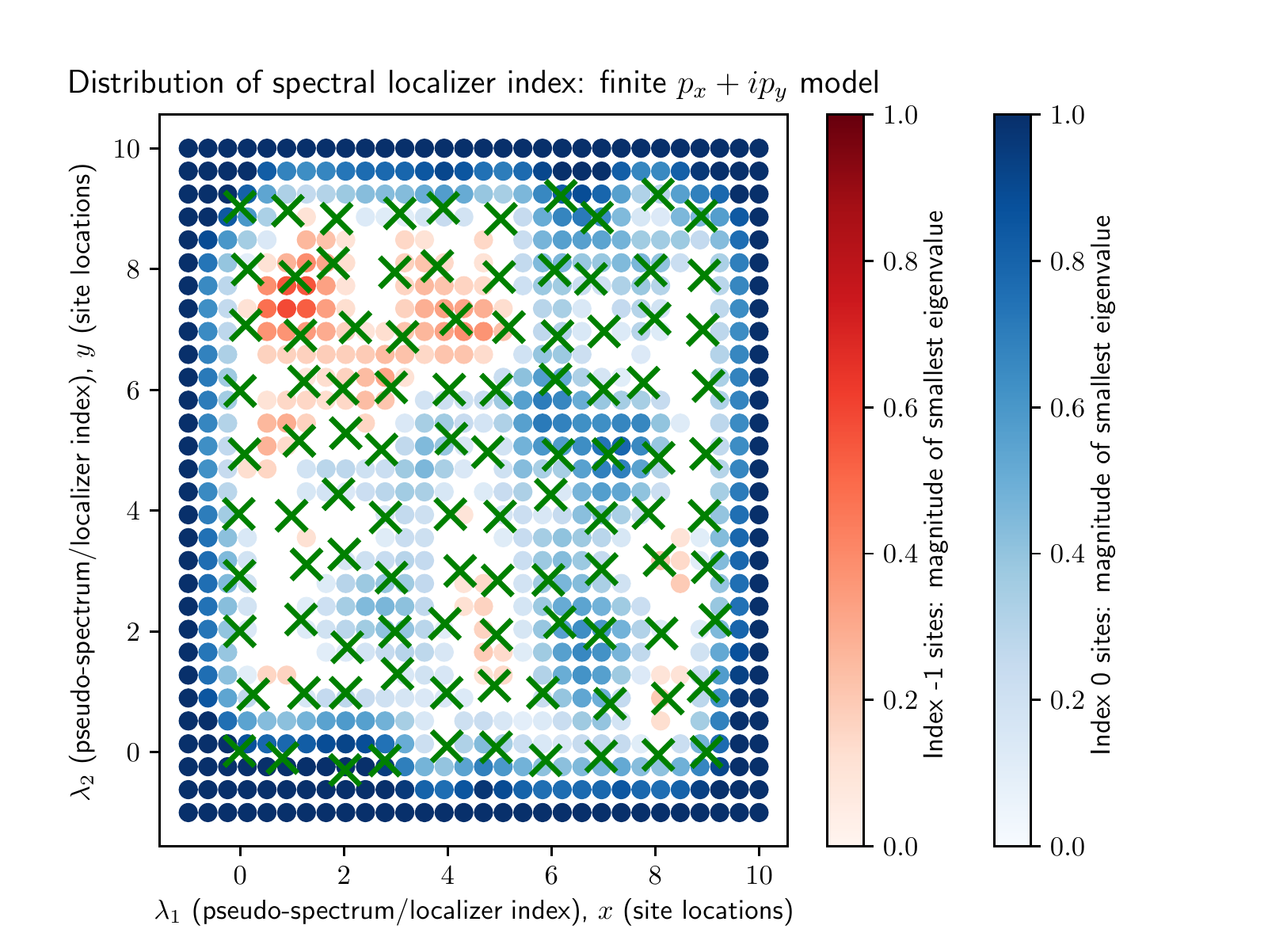}
\caption{} \label{fig:strong_dis_loring}
\end{subfigure}
\begin{subfigure}[b]{\textwidth}
\centering
\includegraphics[scale=.5,draft=false]{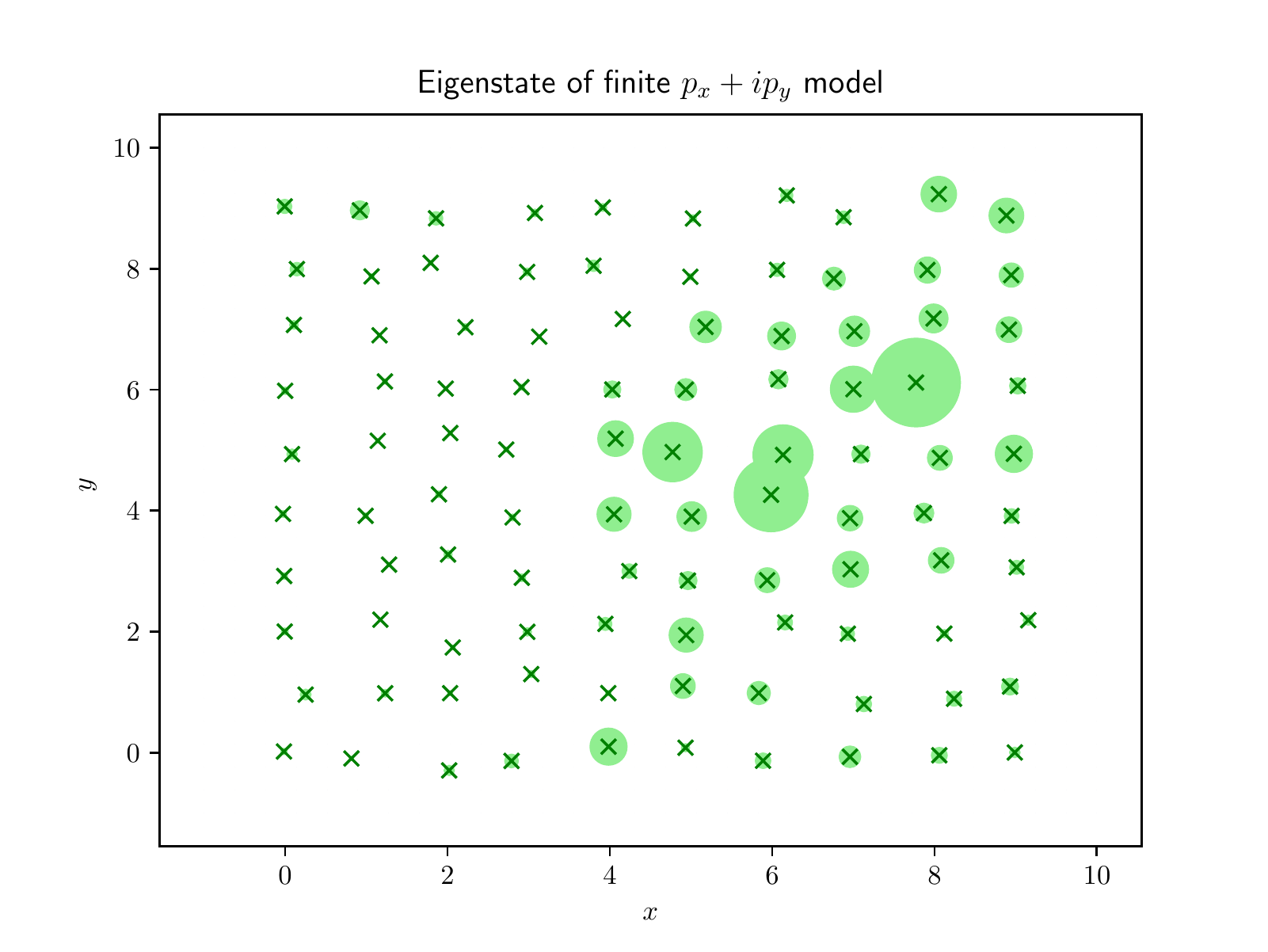}
\caption{} \label{fig:strong_dis_edgestate}
\end{subfigure}
\caption{(a) Computed spectral localizer index and pseudo-spectrum (white) at $\lambda_3 = 0$ for a finite sample modeled by the $p_x + i p_y$ model under strong disorder. For this simulation we took $\mu = 2, \delta = 2, t = 1, \sigma_\mu = 6, \sigma_r = .3$. (b) The absolute value squared of the components of the eigenstate of $H$ with smallest eigenvalue (for the same realization of disorder) at each site, shown by the area of the green circles, with eigenvalue 0.00289 (3sf). The state is no longer supported nearby to the edge of the sample.} 
\label{fig:strong_dis}
\end{figure}

We finally investigated how random disorder affects the distribution of pseudo-spectrum and spectral localizer index. We investigated the effects of three different kinds of disorder: those of potential disorder alone, position disorder alone, and disorder of both kinds. Our results are shown in Figures \ref{fig:pot_dis}, \ref{fig:pos_dis_loring}, and \ref{fig:both_dis}, respectively. In each case, disorder causes the distribution of points where the localizer index is $-1$ to re-shape. When the disorder is strong enough, islands of pseudo-spectrum and index $0$ points can appear within the index $-1$ region. These islands are associated with defect states in the interior of the sample: see Figure \ref{fig:pos_dis_edgestate_3}. We then considered the effect of strong disorder of both kinds, finding that sufficiently strong disorder almost totally destroys the regions of index $-1$: see Figure \ref{fig:strong_dis_loring}. In this case, the eigenstates of $H$ with associated eigenvalues near to $0$ are no longer strongly localized near the edge of the sample (Figure \ref{fig:strong_dis_edgestate}). We note finally that in locations in the interior of the sample where the topological index is unchanged by disorder, the localizer gap tends to decrease significantly in the presence of disorder.

\section{Propagation of wave-packets in $p_x + i p_y$ model} \label{sec:propagation}




\subsection{Finite square lattice with perturbations}

\begin{figure}
\begin{subfigure}[b]{.45\textwidth}
\includegraphics[scale=.35,draft=false]{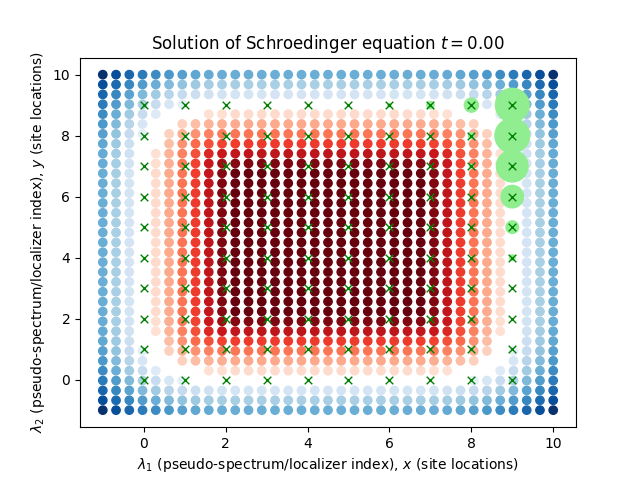}
\end{subfigure}
\begin{subfigure}[b]{.45\textwidth}
\includegraphics[scale=.35,draft=false]{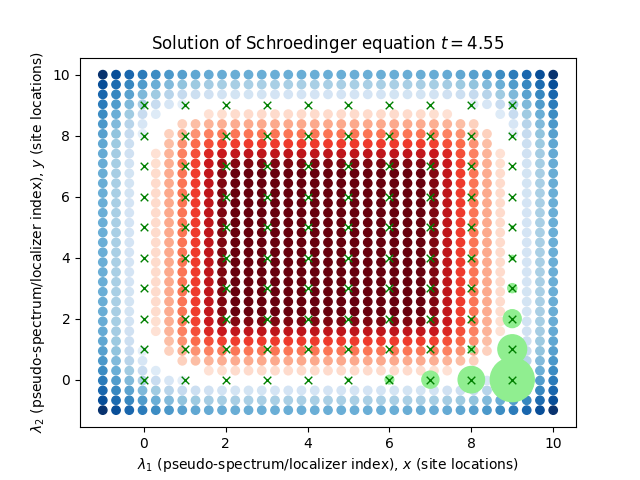}
\end{subfigure}
\begin{subfigure}[b]{.45\textwidth}
\includegraphics[scale=.35,draft=false]{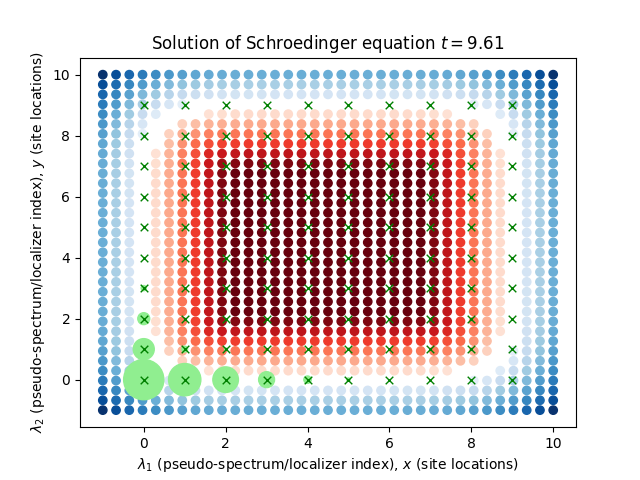}
\end{subfigure}
\begin{subfigure}[b]{.45\textwidth}
\includegraphics[scale=.35,draft=false]{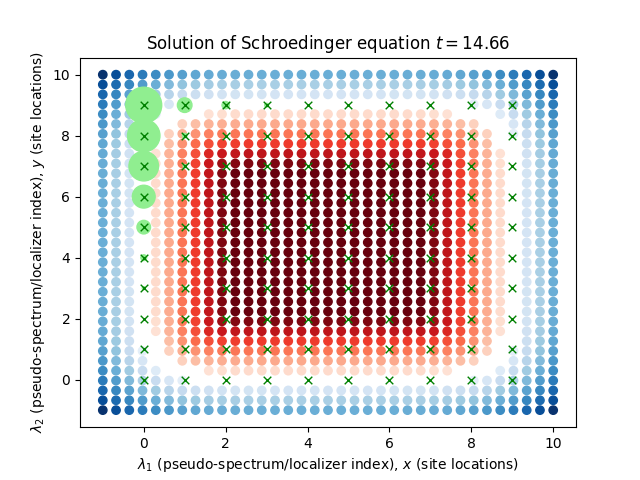}
\end{subfigure}
\begin{subfigure}[b]{.45\textwidth}
\includegraphics[scale=.35,draft=false]{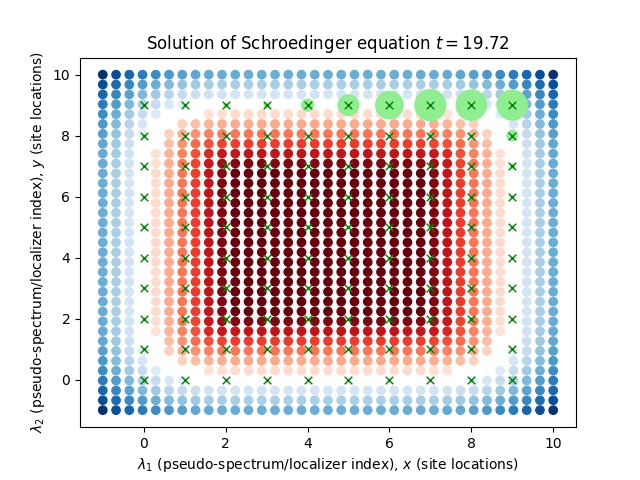}
\end{subfigure}
\begin{subfigure}[b]{.45\textwidth}
\includegraphics[scale=.35,draft=false]{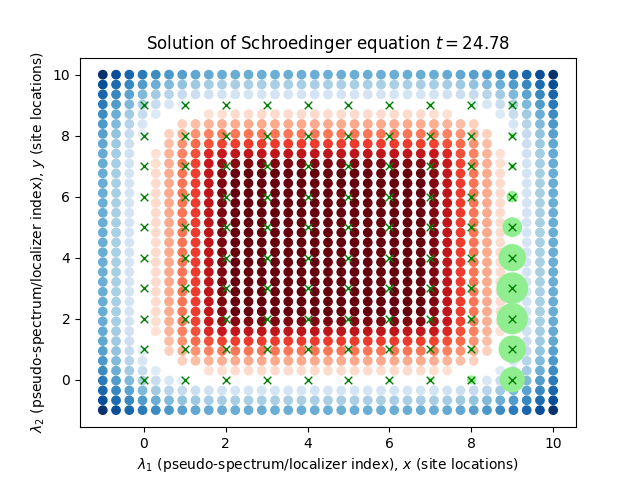}
\end{subfigure}
\begin{subfigure}[b]{.45\textwidth}
\includegraphics[scale=.35,draft=false]{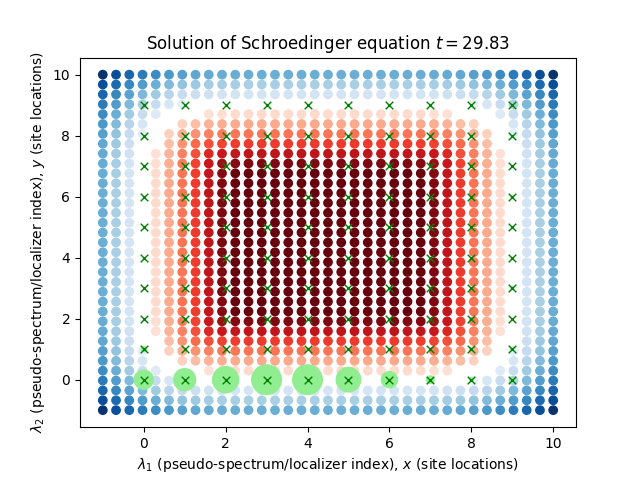}
\end{subfigure}
\begin{subfigure}[b]{.45\textwidth}
\includegraphics[scale=.35,draft=false]{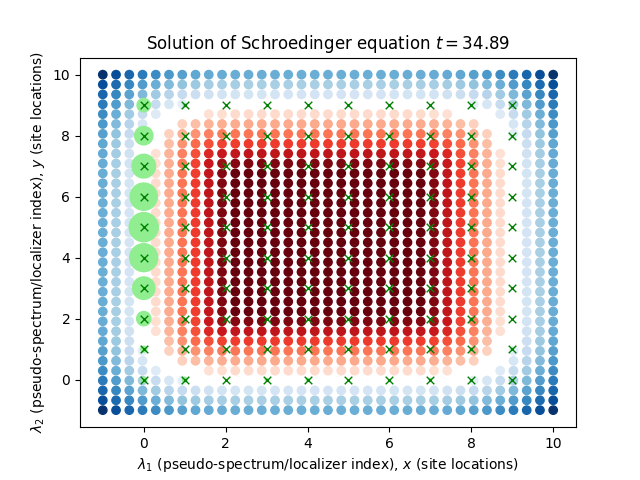}
\end{subfigure}
\caption{In the absence of disorder, a wave-packet built from an edge state of the finite $p_x + i p_y$ model propagates around the edge without back-scattering along the edge or into the bulk. To generate the index information and initial condition for this simulation, we chose $\kappa = .5$. Absolute value squared of components of wave-packet shown by area of green circles.}
\label{fig:prop_nodis}
\end{figure}

We first consider the case without perturbations or disorder, see Figure \ref{fig:prop_nodis}. In this case the wave-packet propagates around the line of pseudospectrum at the edge of the sample with minimal disruption due to back-scattering along the edge or into the bulk. 

\begin{figure}
\begin{subfigure}[b]{.45\textwidth}
\includegraphics[scale=.35,draft=false]{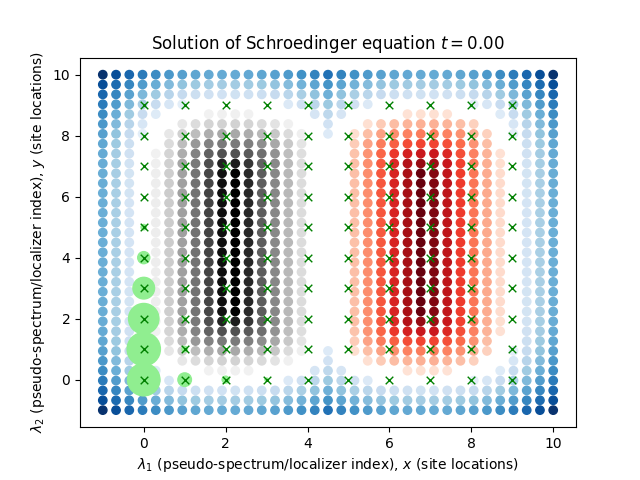}
\end{subfigure}
\begin{subfigure}[b]{.45\textwidth}
\includegraphics[scale=.35,draft=false]{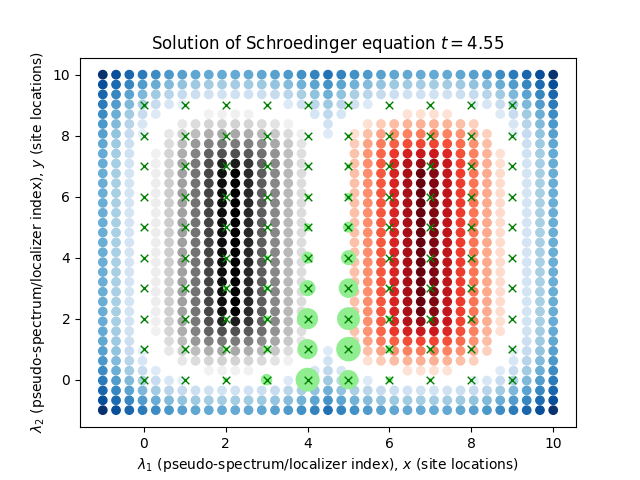}
\end{subfigure}
\begin{subfigure}[b]{.45\textwidth}
\includegraphics[scale=.35,draft=false]{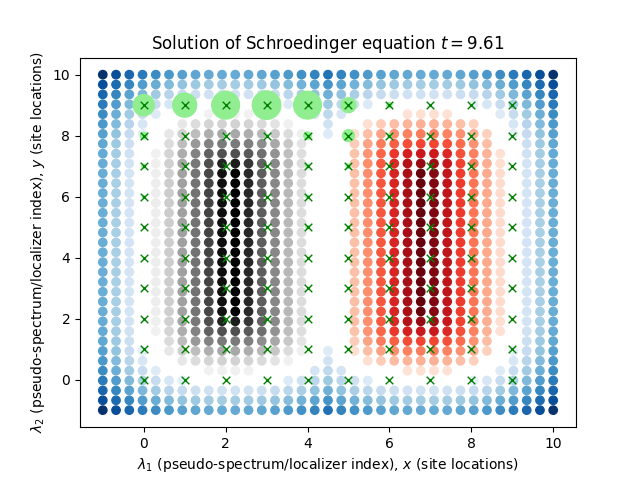}
\end{subfigure}
\begin{subfigure}[b]{.45\textwidth}
\includegraphics[scale=.35,draft=false]{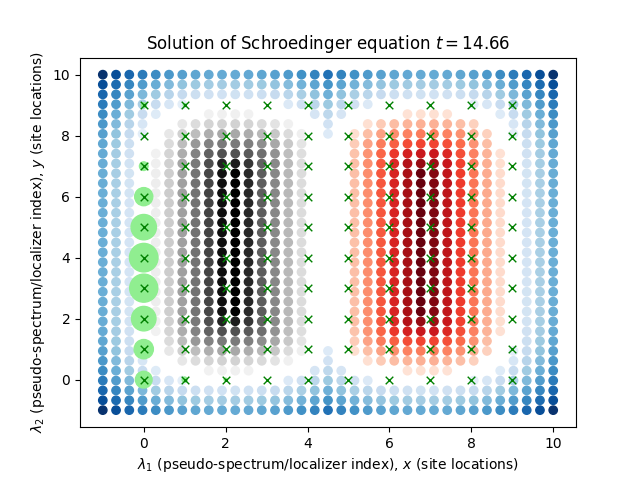}
\end{subfigure}
\begin{subfigure}[b]{.45\textwidth}
\includegraphics[scale=.35,draft=false]{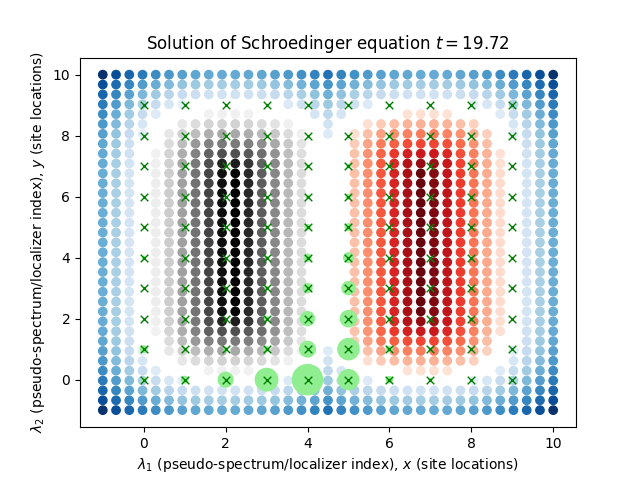}
\end{subfigure}
\begin{subfigure}[b]{.45\textwidth}
\includegraphics[scale=.35,draft=false]{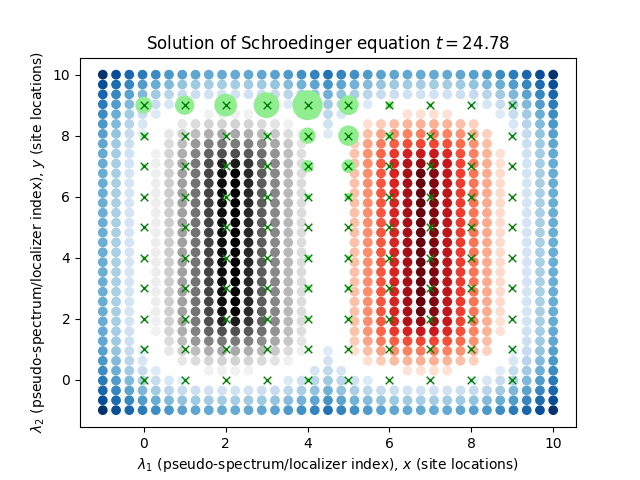}
\end{subfigure}
\begin{subfigure}[b]{.45\textwidth}
\includegraphics[scale=.35,draft=false]{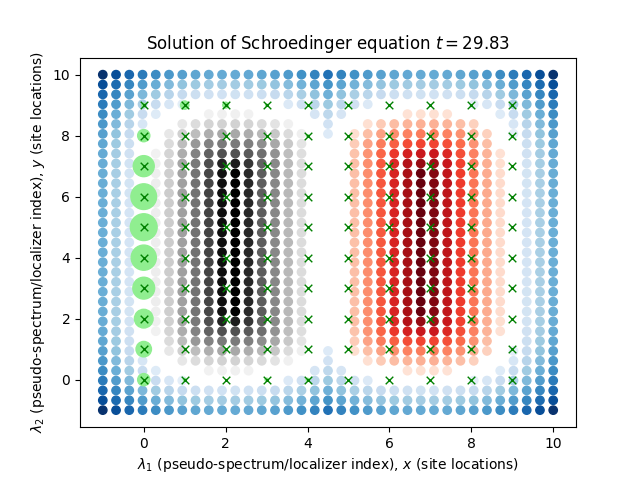}
\end{subfigure}
\begin{subfigure}[b]{.45\textwidth}
\includegraphics[scale=.35,draft=false]{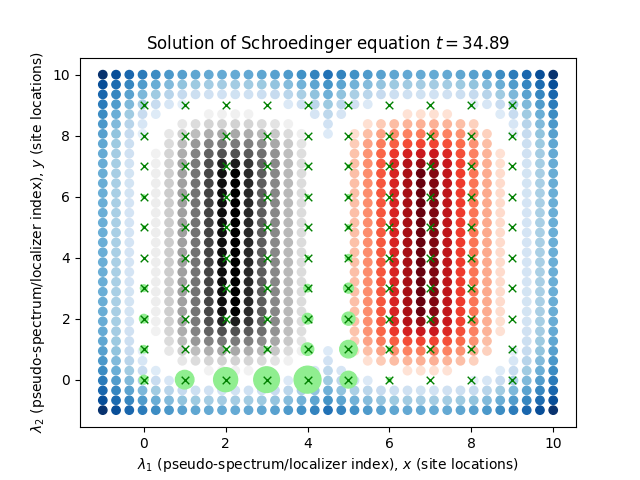}
\end{subfigure}
\caption{We consider the propagation of a localizer state/wave-packet supported at the edge of a region of index $1$, adjacent to a region of index $-1$. We observe that the wave-packet remains localized along the line of pseudospectrum, propagating counter-clockwise around the edge of the index $1$ region. We observe also that the wave-packet does not follow the line of pseudospectrum surrounding the region of index $-1$.}
\label{fig:prop_abrupt}
\end{figure}

\begin{figure}
\begin{subfigure}[b]{.45\textwidth}
\includegraphics[scale=.35,draft=false]{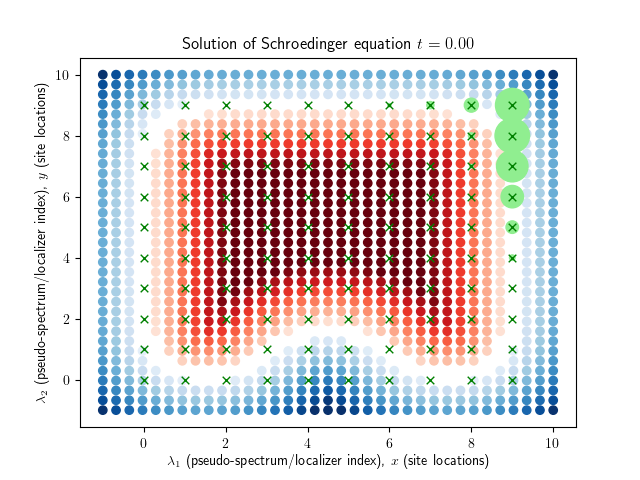}
\end{subfigure}
\begin{subfigure}[b]{.45\textwidth}
\includegraphics[scale=.35,draft=false]{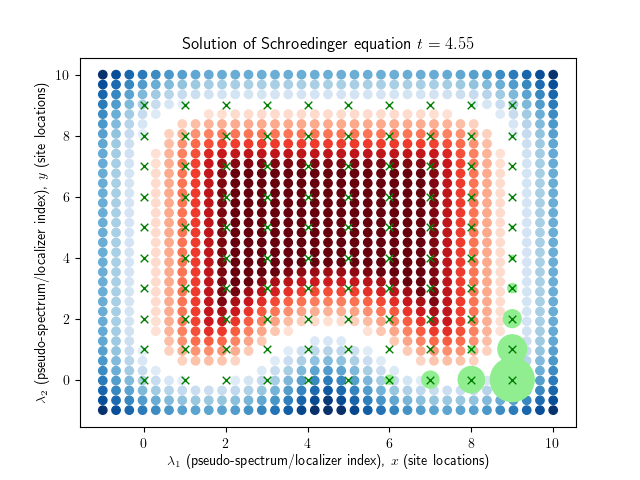}
\end{subfigure}
\begin{subfigure}[b]{.45\textwidth}
\includegraphics[scale=.35,draft=false]{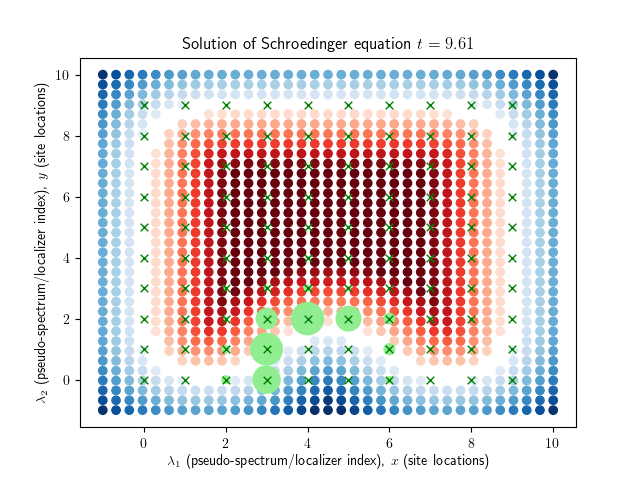}
\end{subfigure}
\begin{subfigure}[b]{.45\textwidth}
\includegraphics[scale=.35,draft=false]{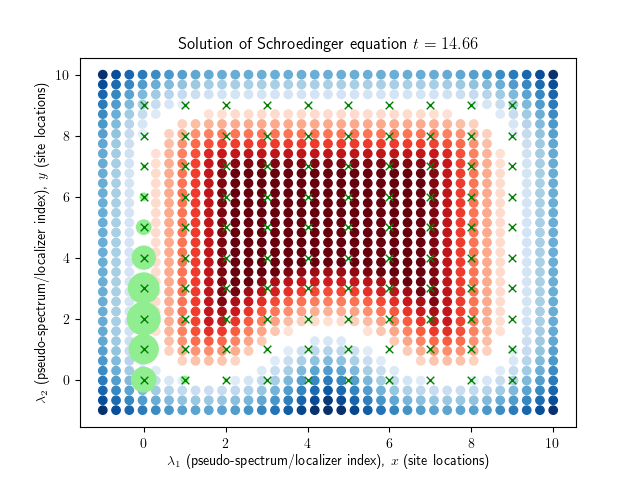}
\end{subfigure}
\begin{subfigure}[b]{.45\textwidth}
\includegraphics[scale=.35,draft=false]{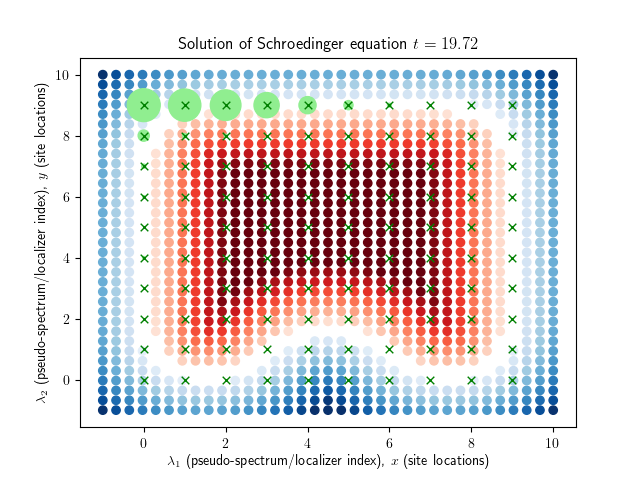}
\end{subfigure}
\begin{subfigure}[b]{.45\textwidth}
\includegraphics[scale=.35,draft=false]{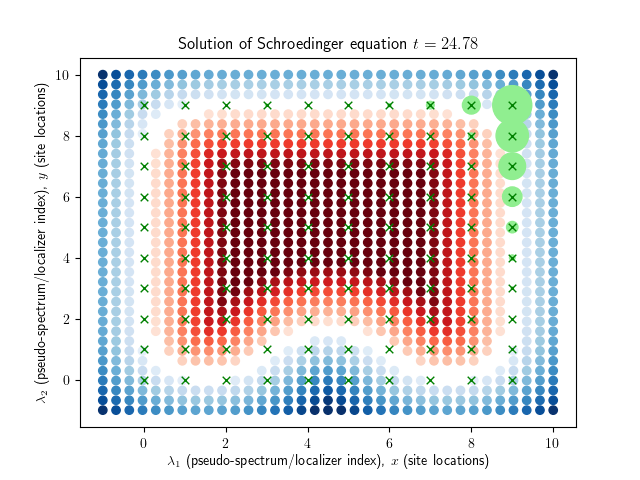}
\end{subfigure}
\begin{subfigure}[b]{.45\textwidth}
\includegraphics[scale=.35,draft=false]{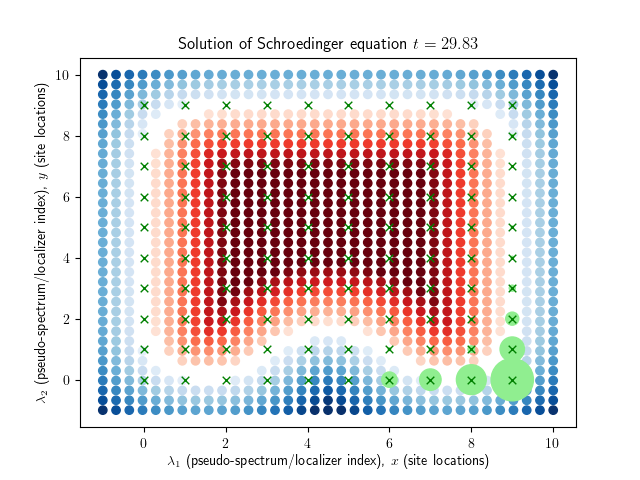}
\end{subfigure}
\begin{subfigure}[b]{.45\textwidth}
\includegraphics[scale=.35,draft=false]{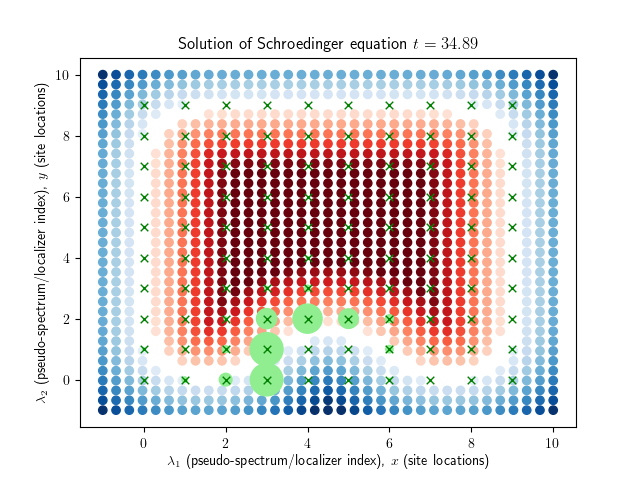}
\end{subfigure}
\caption{We consider the propagation of a localizer state/wave-packet localized at the edge of a region of index $-1$ in the presence of a large, deterministic defect. We observe that the wave-packet remains localized along the line of pseudospectrum at the edge of the structure and does not back-scatter along the edge or into the bulk.}
\label{fig:prop_defect}
\end{figure}

We next considered the case where regions of the structure have different local indices, and studied the dynamics of a wave-packet localized at the edge of one of the regions of non-zero local index. We then considered the case of a wave-packet localized at the edge of a structure where some sites have been effectively removed by adding large onsite potentials. We observed in both cases that the wave-packet remains for $t > 0$ localized to the line of pseudospectrum: see Figures \ref{fig:prop_abrupt} and \ref{fig:prop_defect}. Interestingly, in the case of the structure with regions of different local index (Figure \ref{fig:prop_abrupt}), there is almost no coupling of the wave-packet to the line of pseudo-spectrum surrounding the $-1$ index region.

From these results we draw the following conclusion. Without random disorder, wave-packets initially localized along lines of pseudospectrum remain highly localized to these lines. This conclusion holds even when the structure hosts regions of different index, and when the structure's edge has large defects.

\subsection{Finite square lattice with disorder}

\begin{figure}
\begin{subfigure}[b]{.45\textwidth}
\includegraphics[scale=.35,draft=false]{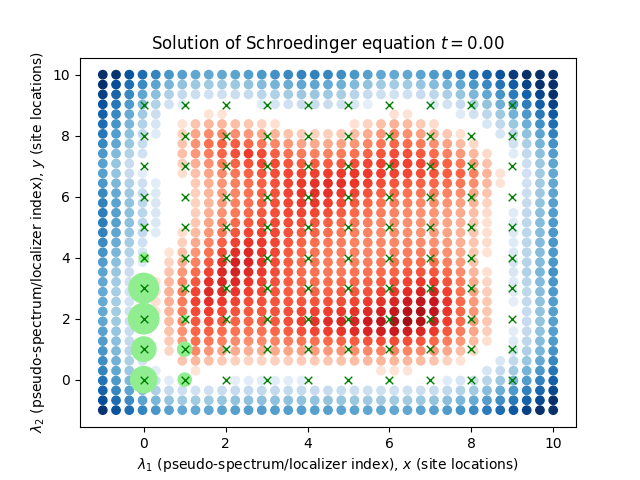}
\end{subfigure}
\begin{subfigure}[b]{.45\textwidth}
\includegraphics[scale=.35,draft=false]{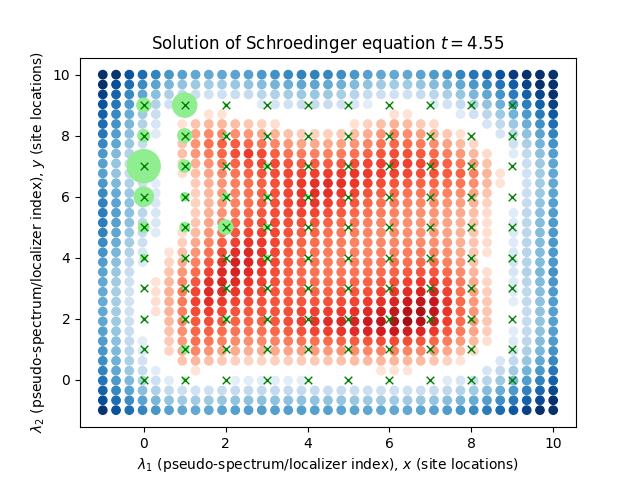}
\end{subfigure}
\begin{subfigure}[b]{.45\textwidth}
\includegraphics[scale=.35,draft=false]{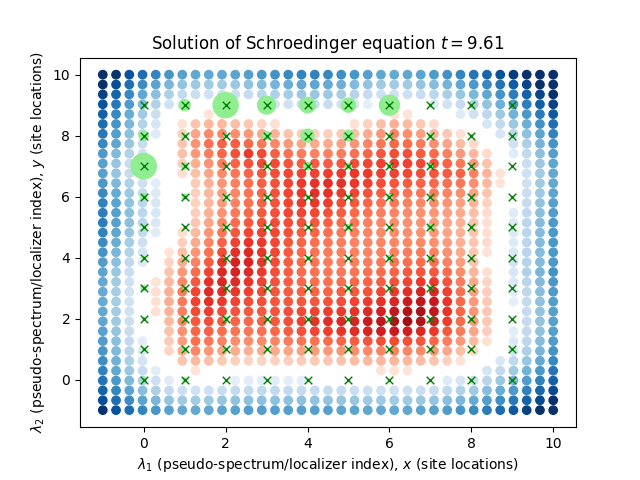}
\end{subfigure}
\begin{subfigure}[b]{.45\textwidth}
\includegraphics[scale=.35,draft=false]{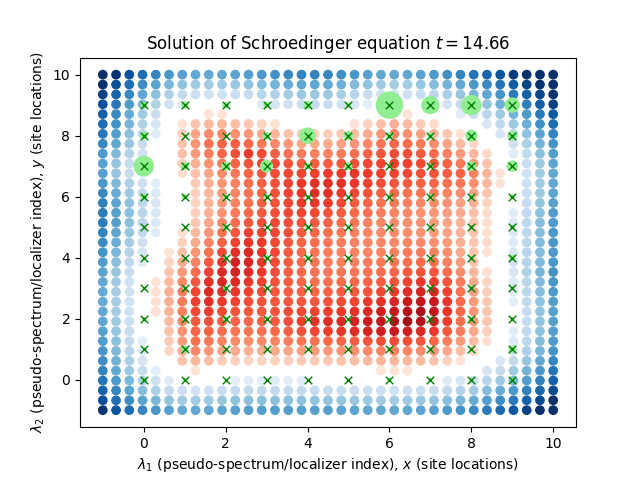}
\end{subfigure}
\begin{subfigure}[b]{.45\textwidth}
\includegraphics[scale=.35,draft=false]{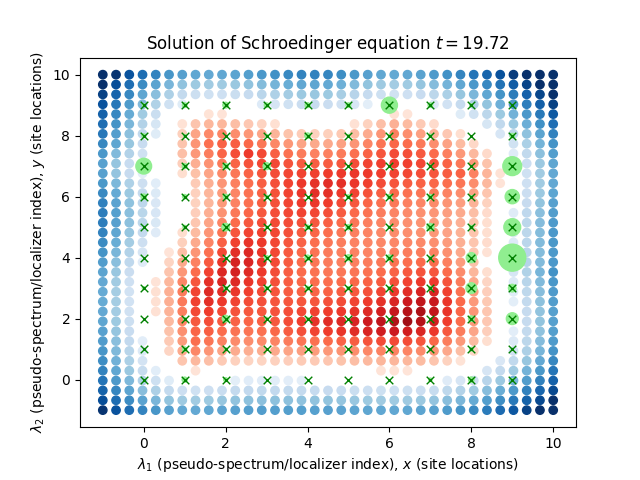}
\end{subfigure}
\begin{subfigure}[b]{.45\textwidth}
\includegraphics[scale=.35,draft=false]{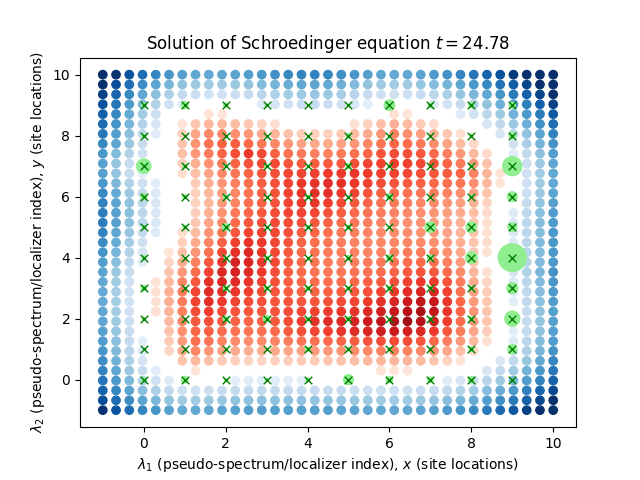}
\end{subfigure}
\begin{subfigure}[b]{.45\textwidth}
\includegraphics[scale=.35,draft=false]{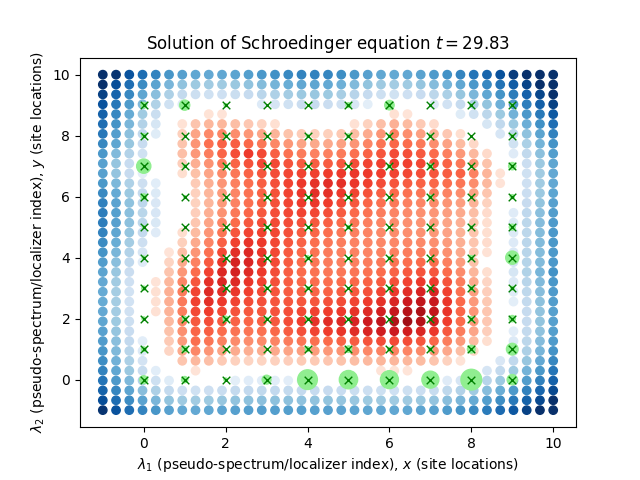}
\end{subfigure}
\begin{subfigure}[b]{.45\textwidth}
\includegraphics[scale=.35,draft=false]{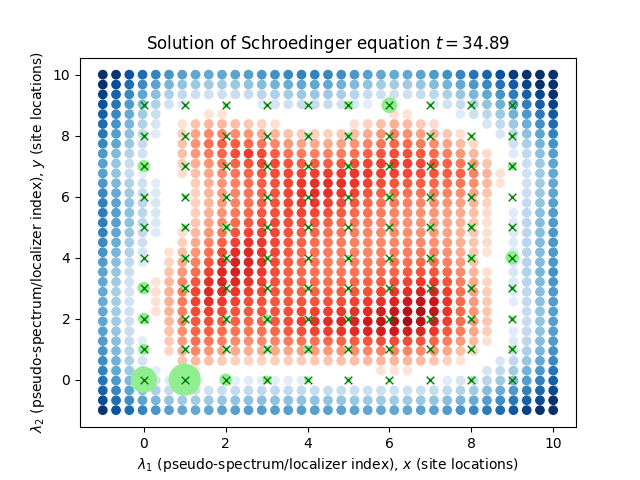}
\end{subfigure}
\caption{In the presence of potential disorder, a localizer state/wave-packet at the edge of the structure propagates mostly without back-scattering, but loses some of its mass to defect states with nearby energy along the edge and in the bulk.
In the figure, the existence of these eigenstates is indicated by regions where the red color of $-1$ index sites is faded (less saturated). At these locations, although the index is $-1$, the localizer has an eigenvalue which is close to zero and hence $H$ has an eigenstate supported at these locations with eigenvalue close to zero.}
\label{fig:prop_potdis}
\end{figure}

\begin{figure}
\begin{subfigure}[b]{.45\textwidth}
\includegraphics[scale=.35,draft=false]{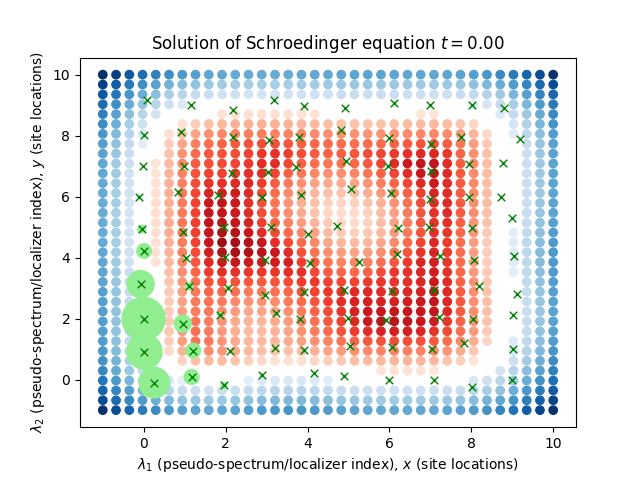}
\end{subfigure}
\begin{subfigure}[b]{.45\textwidth}
\includegraphics[scale=.35,draft=false]{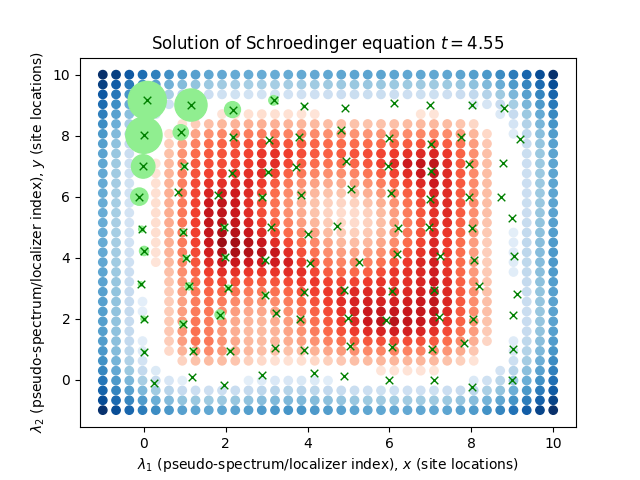}
\end{subfigure}
\begin{subfigure}[b]{.45\textwidth}
\includegraphics[scale=.35,draft=false]{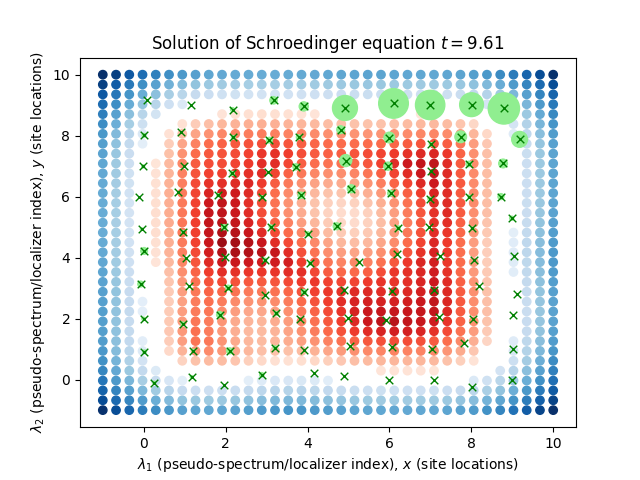}
\end{subfigure}
\begin{subfigure}[b]{.45\textwidth}
\includegraphics[scale=.35,draft=false]{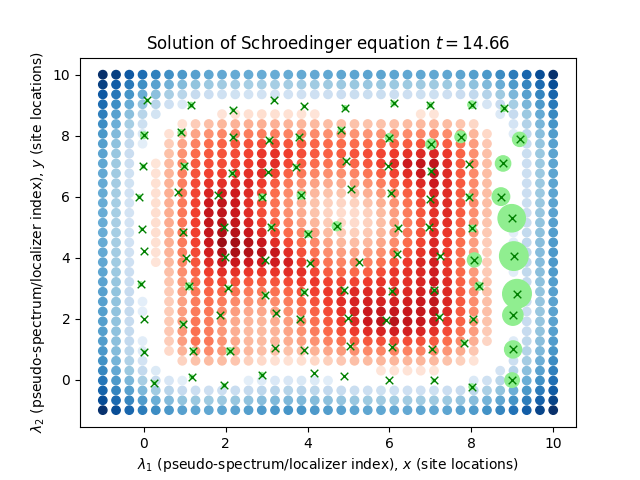}
\end{subfigure}
\begin{subfigure}[b]{.45\textwidth}
\includegraphics[scale=.35,draft=false]{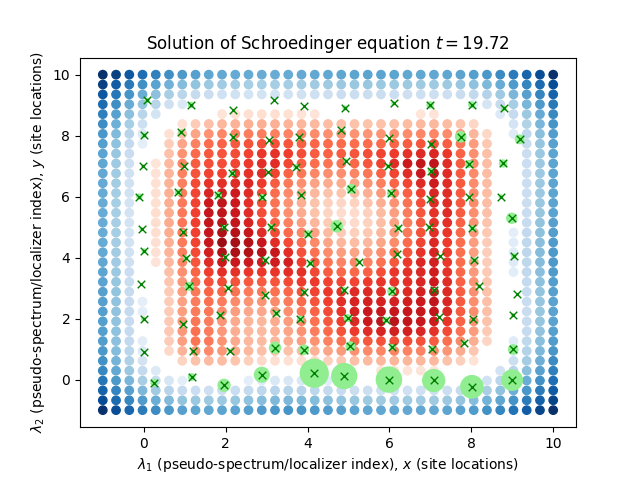}
\end{subfigure}
\begin{subfigure}[b]{.45\textwidth}
\includegraphics[scale=.35,draft=false]{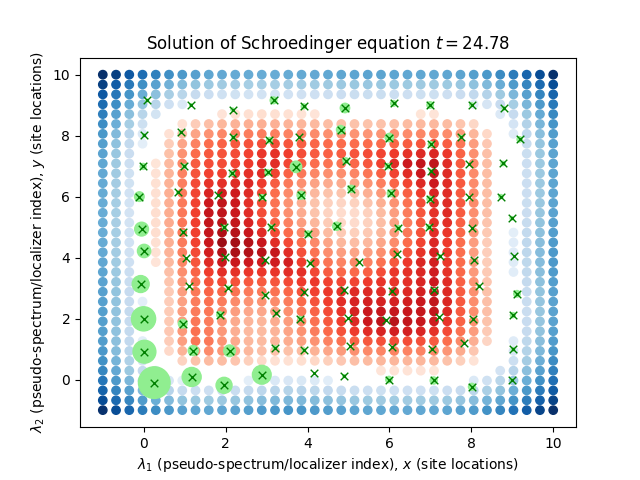}
\end{subfigure}
\begin{subfigure}[b]{.45\textwidth}
\includegraphics[scale=.35,draft=false]{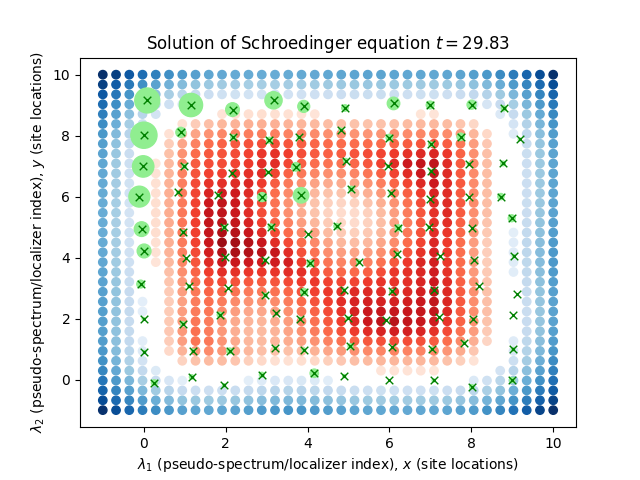}
\end{subfigure}
\begin{subfigure}[b]{.45\textwidth}
\includegraphics[scale=.35,draft=false]{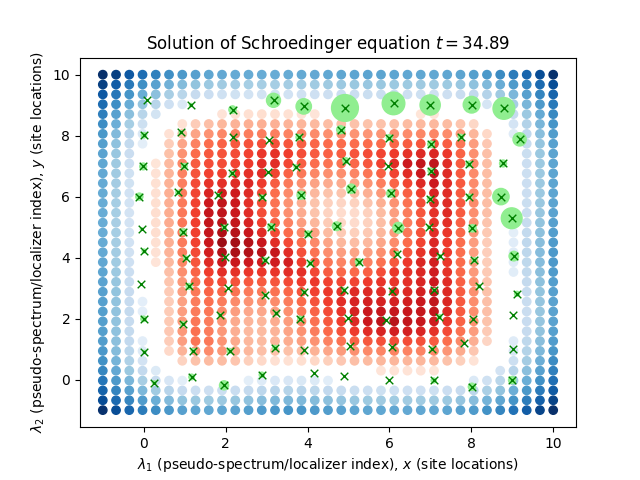}
\end{subfigure}
\caption{In the presence of position disorder, a wave-packet at the edge of the structure propagates mostly without back-scattering, but loses some of its mass to defect states with nearby energy along the edge and in the bulk.}
\label{fig:prop_posdis}
\end{figure}

\begin{figure}
\begin{subfigure}[b]{.45\textwidth}
\includegraphics[scale=.35,draft=false]{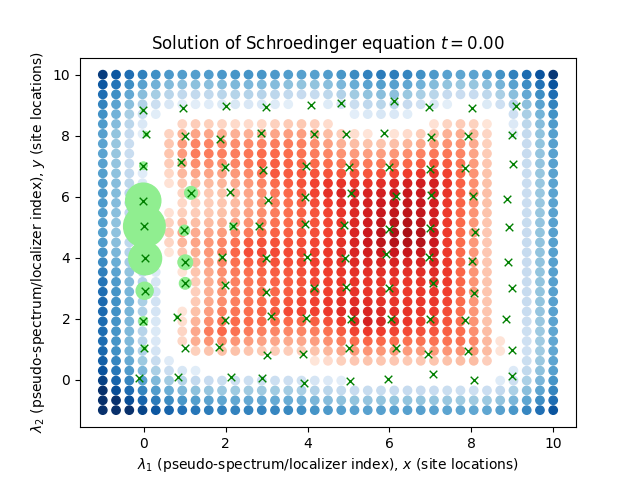}
\end{subfigure}
\begin{subfigure}[b]{.45\textwidth}
\includegraphics[scale=.35,draft=false]{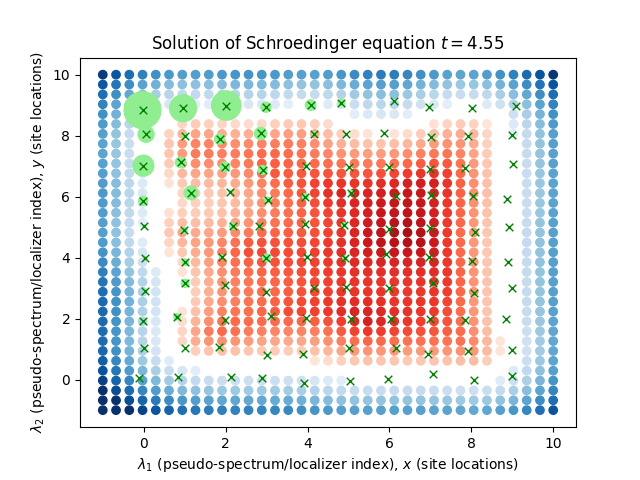}
\end{subfigure}
\begin{subfigure}[b]{.45\textwidth}
\includegraphics[scale=.35,draft=false]{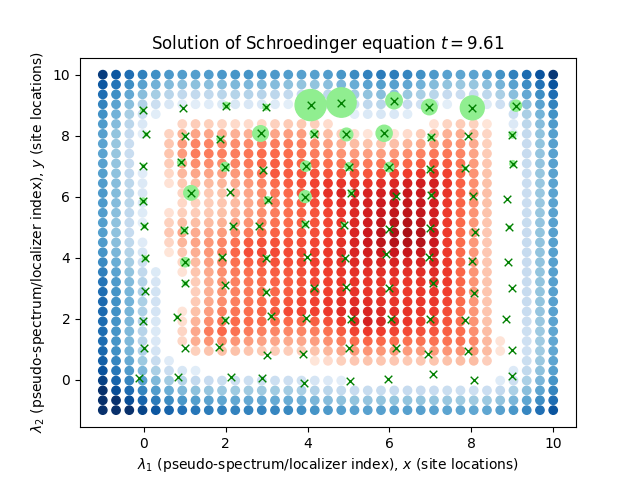}
\end{subfigure}
\begin{subfigure}[b]{.45\textwidth}
\includegraphics[scale=.35,draft=false]{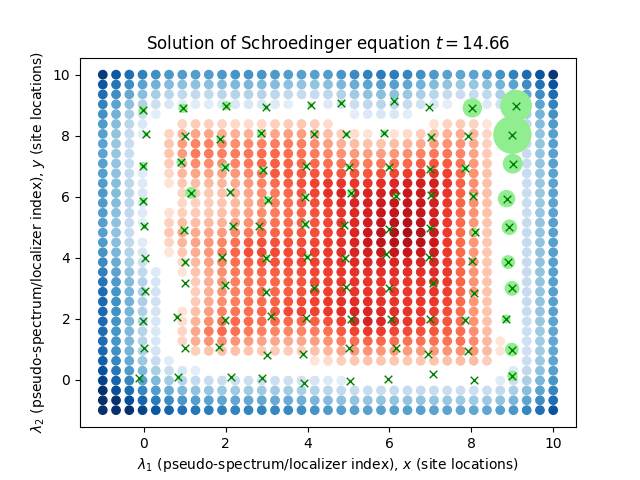}
\end{subfigure}
\begin{subfigure}[b]{.45\textwidth}
\includegraphics[scale=.35,draft=false]{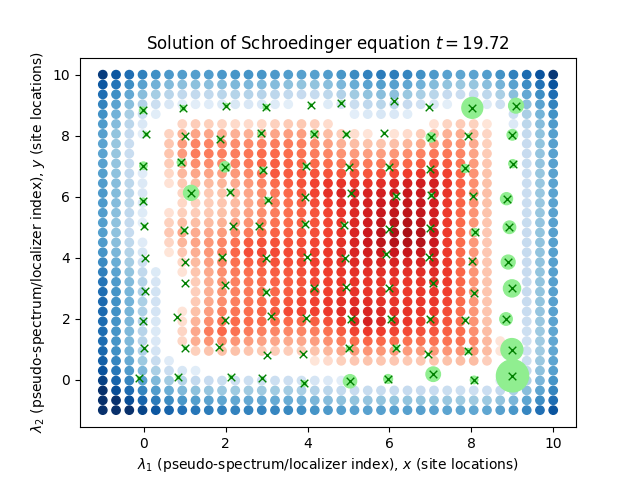}
\end{subfigure}
\begin{subfigure}[b]{.45\textwidth}
\includegraphics[scale=.35,draft=false]{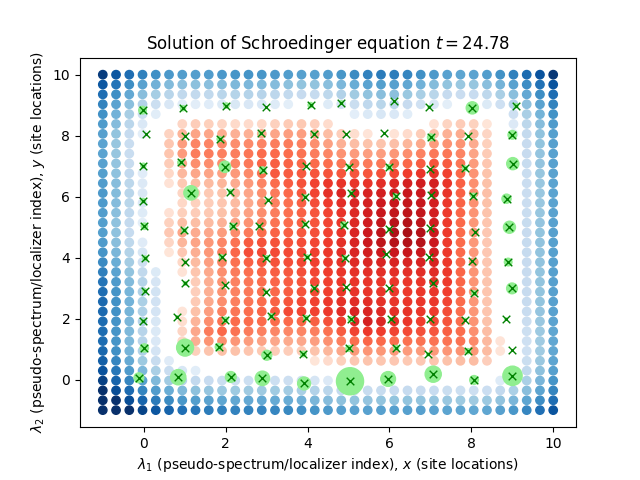}
\end{subfigure}
\begin{subfigure}[b]{.45\textwidth}
\includegraphics[scale=.35,draft=false]{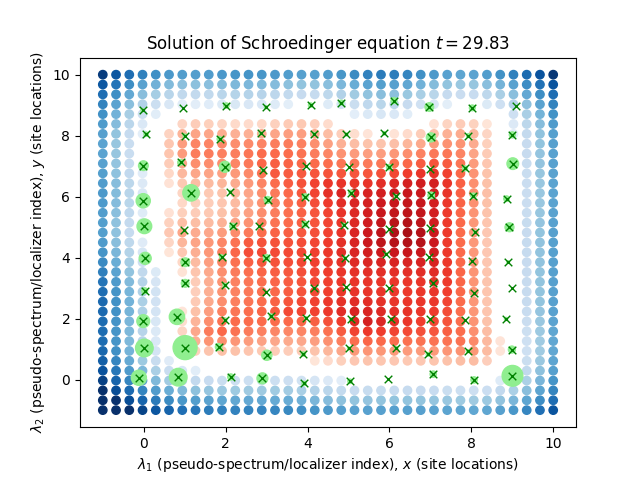}
\end{subfigure}
\begin{subfigure}[b]{.45\textwidth}
\includegraphics[scale=.35,draft=false]{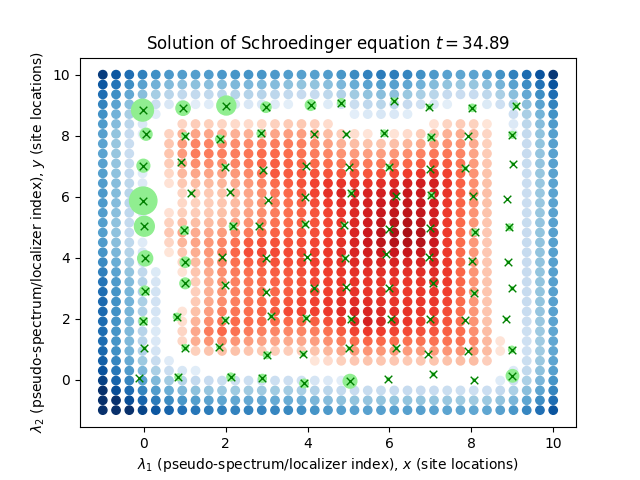}
\end{subfigure}
\caption{In the presence of both kinds of disorder, a wave-packet built from an edge state of the finite $p_x + i p_y$ model propagates around the edge mostly without back-scattering, but loses some of its mass to defect states with nearby energy along the edge and in the bulk.}
\label{fig:prop_bothdis}
\end{figure}

\begin{figure}
\begin{subfigure}[b]{.45\textwidth}
\includegraphics[scale=.35,draft=false]{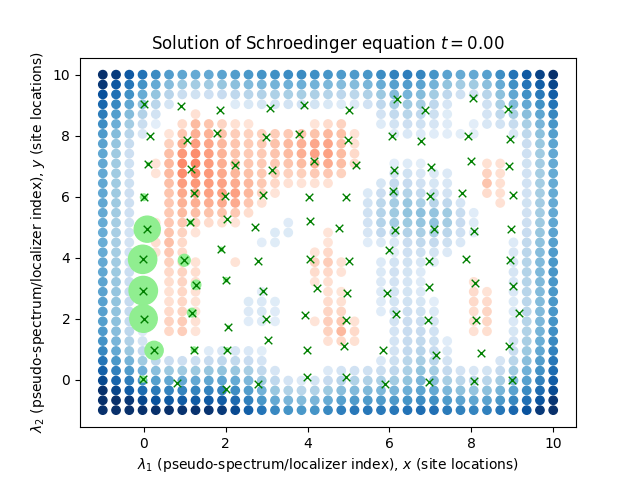}
\end{subfigure}
\begin{subfigure}[b]{.45\textwidth}
\includegraphics[scale=.35,draft=false]{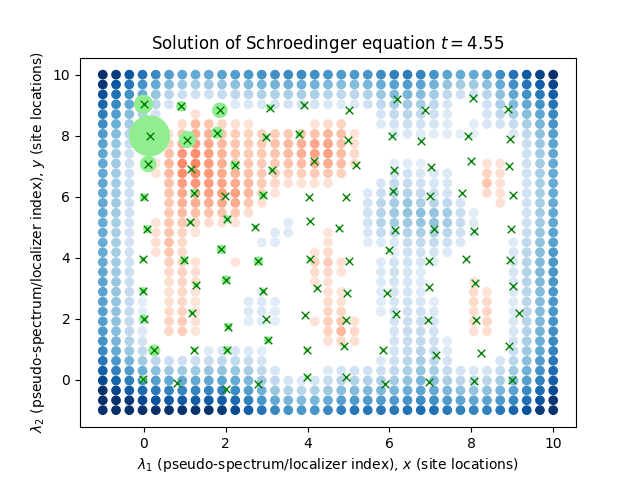}
\end{subfigure}
\begin{subfigure}[b]{.45\textwidth}
\includegraphics[scale=.35,draft=false]{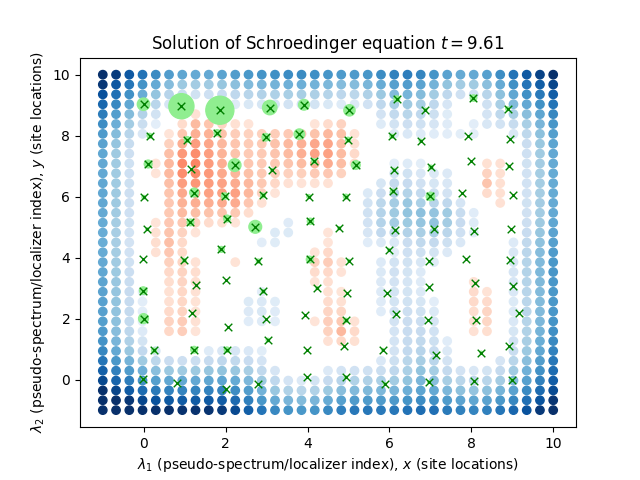}
\end{subfigure}
\begin{subfigure}[b]{.45\textwidth}
\includegraphics[scale=.35,draft=false]{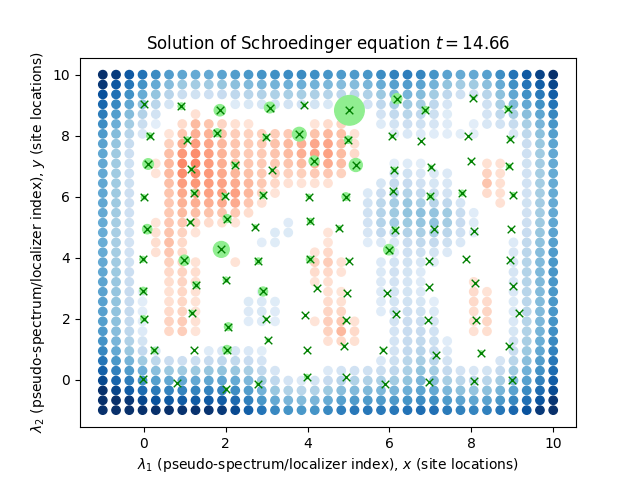}
\end{subfigure}
\begin{subfigure}[b]{.45\textwidth}
\includegraphics[scale=.35,draft=false]{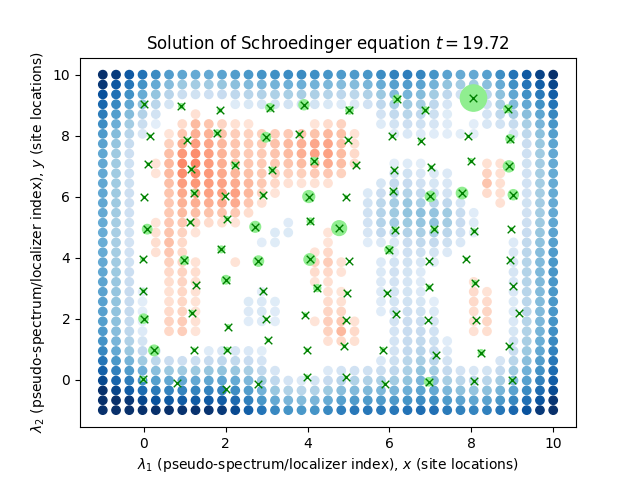}
\end{subfigure}
\begin{subfigure}[b]{.45\textwidth}
\includegraphics[scale=.35,draft=false]{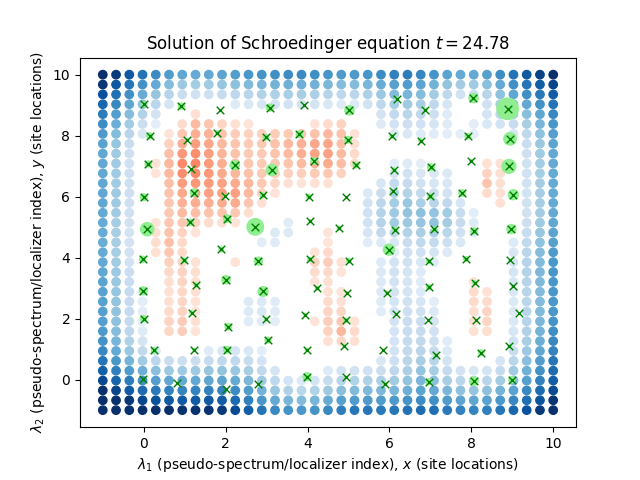}
\end{subfigure}
\begin{subfigure}[b]{.45\textwidth}
\includegraphics[scale=.35,draft=false]{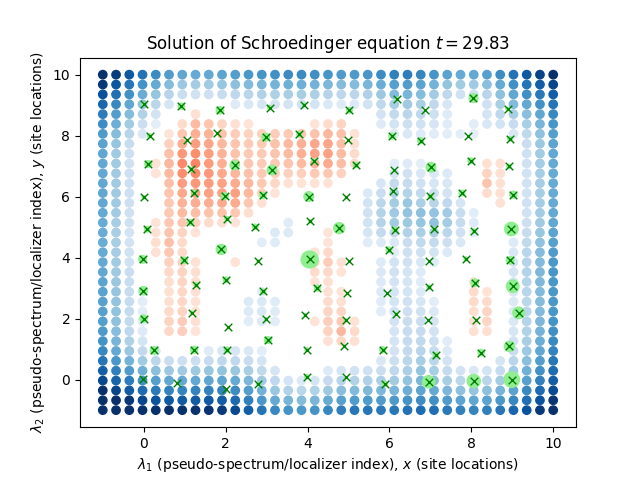}
\end{subfigure}
\begin{subfigure}[b]{.45\textwidth}
\includegraphics[scale=.35,draft=false]{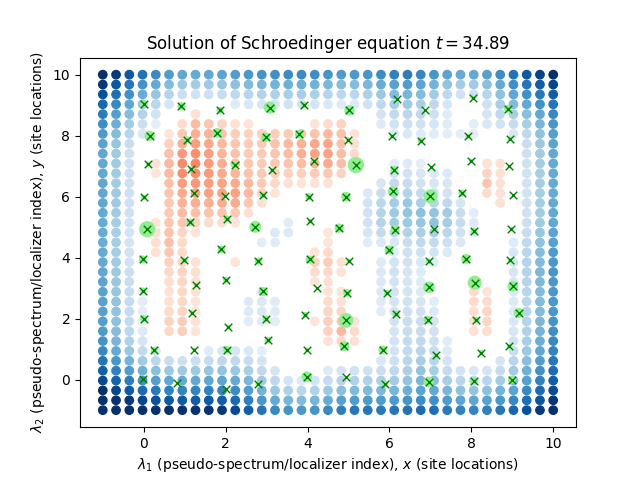}
\end{subfigure}
\caption{Propagation of a wave-packet at the edge of the structure in the presence of strong disorder. The solution disperses almost completely throughout the structure.}
\label{fig:prop_strongdis}
\end{figure}

We then considered wave-packet propagation in the presence of varying amounts of disorder: potential disorder only (Figure \ref{fig:prop_potdis}), position disorder only (Figure \ref{fig:prop_posdis}), both kinds of disorder (Figure \ref{fig:prop_bothdis}), and strong disorder (Figure \ref{fig:prop_strongdis}). We observe that in the presence of disorder, wave-packets tend to spread into the bulk as they propagate. We interpret this as follows. With disorder, the Hamiltonian $H$ can have eigenstates supported in the interior of the structure with associated eigenvalues which are close to zero. The spreading of the wave-packet into the bulk can be understood as resulting from the wave-packet coupling to these eigenstates as it propagates. 

The existence of eigenstates of $H$ with near-zero eigenvalues in the presence of disorder is indicated in the figures generated in the previous section by the lighter color of parts of the region of index $-1$. Recall that the color saturation of points outside the pseudospectrum is defined by the localizer gap, i.e. the magnitude of the eigenvalue of the localizer closest to $0$, with lighter saturations corresponding to a smaller gap. Whenever $H$ has an eigenstate with eigenvalue near to zero supported within the region of index $-1$, the localizer will have a small magnitude eigenvalue at $\lambda_1, \lambda_2$, and hence the localizer gap will shrink there when $\lambda_3 = 0$. 

From these results we note the following connection between the plots of localizer index we saw in Section \ref{sec:ind_comp} and the wave-packet dynamics shown in Figures \ref{fig:prop_potdis}-\ref{fig:prop_strongdis}. Whenever the localized index is weak adjacent to the curve of pseudo-spectrum, wave-packets no longer propagate robustly along the curve.
Note that in all of the examples we considered without disorder, although the regions of localizer index and curves of pseudo-spectrum re-shaped when we introduced defects, the localizer index remained strong on either side of the pseudo-spectrum curve. 
We leave the problem of understanding, in a more quantitative and precise way, this apparent link between localizer index strength and wave-packet propagation to future work. 


\section{Conclusions and perspectives} \label{sec:fut}

\subsection{Conclusions}

In this work we have studied the spectral localizer index and to what extent it predicts propagation of wave-packets in a two-dimensional finite topological insulator. Our simulations suggest that whenever the localizer gap is sufficiently large (in the sense of Sections \ref{sec:strength} and \ref{sec:strength_2}) adjacent to a curve of pseudo-spectrum, wave-packets (more precisely, initial data given by localizer states) propagate robustly along this curve. Disorder has the effect of weakening the strength of the localizer index, and hence wave-packets may no longer propagate robustly in this case. 

\subsection{Perspectives}

To our knowledge, the present work is the first to propose any connection between the localizer and solutions of the \emph{time-dependent} Schr\"odinger equation \eqref{eq:schro}. As a long-term goal, it would be very interesting if the connection could be made quantitative, and then verified analytically.

In the meantime, we plan to further test the conclusions of the present study. First, we would like to test what happens for larger system sizes. Second, we would like to test what happens with different choices of tuning for the localizer. We could experiment, for example, with choosing different values of the tuning parameter $\kappa$ to generate plots of the localizer pseudospectrum and index and to generate localizer state initial conditions.




\printbibliography





\end{document}